\pdfoutput=1
\documentclass[final,5p,times,twocolumn]{elsarticle}
\usepackage{amsmath,amsfonts}
\usepackage{algorithmic}
\usepackage{algorithm}
\usepackage{array}
\usepackage[caption=false,font=footnotesize,labelfont=rm,textfont=rm]{subfig}
\usepackage{textcomp}
\usepackage{url}
\usepackage{verbatim}
\usepackage{graphicx}

\usepackage{xcolor} 
\usepackage{tikz} 
\usepackage{enumitem} 
\usepackage{float} 
\usetikzlibrary{arrows,chains,positioning,scopes,shapes.geometric,shapes.misc,shadows} 
\usepackage[capitalise]{cleveref} 
\crefformat{appendix}{#2#1#3}\Crefformat{appendix}{#2#1#3}
\crefformat{subappendix}{#2#1#3}\Crefformat{subappendix}{#2#1#3}

\journal{Fusion Engineering and Design}
\biboptions{sort&compress} 

\begin{document}

\begin{frontmatter}

\title{Coordinating integration of multidisciplinary design analysis and optimisation codes with design structure matrices}

\author[tue-ap]{W.J. Rutten\corref{cor1}}
\ead{w.j.rutten@tue.nl}
\author[tue-me]{L.F.P. Etman}
\author[ipp,tue-ap]{F. Warmer}
\cortext[cor1]{Corresponding author}

\affiliation[tue-ap]{organization={Department of Applied Physics and Science Education, Eindhoven University of Technology},
            city={Eindhoven},
            country={The Netherlands}}
\affiliation[tue-me]{organization={Department of Mechanical Engineering, Eindhoven University of Technology},
            city={Eindhoven},
            country={The Netherlands}}
\affiliation[ipp]{organization={Max Planck Institute for Plasma Physics},
            city={Greifswald},
            country={Germany}}

\begin{abstract}
The integrated analysis and optimisation codes used to design nuclear fusion reactors need to become increasingly large scale, multidisciplinary and complex to computationally verify design requirements more thoroughly and earlier in the design process.
Though the manner in which the integrated models are coupled can have significant implications for computational efficiency, the structure of couplings typically remains hidden in the implementation of a design framework.
As this problem is not unique to fusion, we frame integrated design problems as applications of the broader field of Multidisciplinary Design Analysis and Optimisation (MDAO).
Originally developed for complex system design problems in e.g. aerospace, the MDAO field uses (extended) Design Structure Matrices (DSMs) to display the architecture of integrated analysis and optimisation codes.

The aim of this work is to introduce these DSM methods and demonstrate how they can be used to improve the transparency and efficiency of design codes in fusion.
As a proof of concept, DSM methods are applied to two case studies: stellarator optimisation with SIMSOPT and the PROCESS systems code.
With these cases, we illustrate how a DSM can help identify structural assumptions in the code, and how architectural insight can be used to improve computational efficiency.
In particular, we demonstrate the computational impact by modifying the PROCESS solver architecture based on the feedback loops revealed in an automatically generated DSM, yielding a significant reduction in model evaluations.
Such architectural changes are considered especially impactful for future design frameworks that combine higher fidelity models with large numbers of inexpensive models.
\end{abstract}

\begin{keyword}
Systems Codes \sep Stellarator Optimisation \sep Integrated Modelling \sep Design Structure Matrix \sep xDSM \sep Multidisciplinary Design Analysis and Optimisation 
\end{keyword}

\end{frontmatter}

\section{Introduction}
\label{sec:introduction}
Optimisation is commonly used in the conceptual design of fusion reactors to explore the design space and find suitable candidate concepts.
Systems codes optimise a complete fusion reactor using simple models to represent all physics and engineering disciplines~\cite{morrisPreparingSystemsCodes2021, muldrewConceptualDesignWorkflow2024}.
This yields self-consistent and optimal concepts to be used as a starting point for detailed analysis of the specific disciplines.
Such integrated analysis and optimisation codes inform decision making in the early design phase, and in turn have a significant impact on the success of the design project~\cite{brevaultAerospaceSystemAnalysis2020}.

Enabled by the increasing availability of computational power, this model-based engineering approach is expanding to verify the feasibility of requirements more thoroughly in earlier design phases.
This is expected to reduce development cost, time and risk as fewer prototypes and experiments are required, and more alternative designs can be considered in more detail~\cite{pattersonIntegratedDigitalFramework2019, duganUsingVirtualEngineering2021}. 
Moreover, key requirements (e.g.\ regarding nuclear safety and fusion performance) may necessitate high fidelity verification \textit{in silico}, as physical experiments are expensive or even impossible~\cite{kemptonMaturingDesignChallenges2024}. 
Eventually, such approaches are envisioned to develop into digital twins which support not only reactor design but also operation with multi-fidelity integrated models~\cite{battyeDigitalTwinsFusion2025,jenkoDigitalTwinsFusion2026a}.

A key aspect of such approaches is the modularity of analysis models in an overarching framework, which enables the codes to keep the software structured when integrating more disciplines.
Various modular frameworks have been developed for integrated design analysis (e.g.\ JINTRAC~\cite{romanelliJINTRACSystemCodes2014a}, OMFIT~\cite{meneghiniIntegratedModelingApplications2015}, FERMI~\cite{badalassiFERMIFusionEnergy2023a}), often focusing on specific parts of a fusion reactor~\cite{smolentsevRoleIntegratedComputer2020}.
Reactor design optimisation codes such as PROCESS~\cite{kovariPROCESSSystemsCode2014,kovariPROCESSSystemsCode2016}, SYCOMORE~\cite{reuxDEMOReactorDesign2015a} and FUSE~\cite{meneghiniFUSEFusionSynthesis2024a} also feature various degrees of modularity, but are limited to lower fidelity models.
Initial approaches to integrate higher fidelity models include the CFETR Integration Design Platform~\cite{yeDevelopmentCFETRIntegration2017,maoCFETRIntegrationDesign2019} and BLUEMIRA~\cite{colemanBLUEMIRAModularOpensource2026a}, which combines the BLUEPRINT~\cite{colemanBLUEPRINTNovelApproach2019} and MIRA~\cite{franzaMIRAMultiphysicsApproach2022} codes.
Such approaches are particularly important for stellarator design, as 3D shape optimisation (with e.g.\ SIMSOPT~\cite{landremanSIMSOPTFlexibleFramework2021}, DESC~\cite{dudtDESCStellaratorCode2023}) lies at the heart of a stellarator concept~\cite{lionGeneralStellaratorVersion2021a}.
Enabling integration of systems codes and stellarator optimisation approaches is the direct motivation for this work~\cite{warmerOverviewEuropeanEfforts2024a}.

Even with a modular approach, the increasing scale and multidisciplinarity of design codes come at the cost of transparency, as it becomes difficult to keep an overview of the complex dependencies between analysis models. 
This challenges validation of the integrated model, as it is hard to assess if all relevant interfaces are captured, and if not, what assumptions have been made.
Especially for cross-disciplinary dependencies this is critical, as they may contain interface assumptions that are not easily caught by disciplinary experts.
This opacity subsequently affects the interpretability of the results, as it is difficult to determine what models and assumptions have led to the produced results.

Optimisation problems concerning large-scale, complex, multidisciplinary systems are not unique to fusion.
Originating in aerospace, the field of Multidisciplinary Design Analysis and Optimisation (MDAO) has emerged to address such problems~\cite{alexandrovMultidisciplinaryDesignOptimization1997, sobieszczanski-sobieskiMultidisciplinaryDesignOptimization2015,martinsEngineeringDesignOptimization2021}. 
Here multidisciplinary refers to the modularity of the disciplinary codes that are integrated: large-scale codes typically remain modular and are coupled via strictly defined interfaces.
For complex systems, the resulting couplings tend to create feedback between models that cannot be decoupled without sacrificing accuracy of results.
This approach is not limited to design optimisation codes; it generally applies to integrated modelling approaches as well, in which case it is referred to as Multidisciplinary Analysis (MDA).
Though the choice of disciplinary models and algorithms is important, the MDAO field focuses in particular on the arrangement of models, solvers and optimisers: the MDAO architecture. 
Many MDAO architectures can be defined to solve the same mathematical optimisation problem, structuring the couplings between models and drivers in different ways~\cite{tosseramsClassificationMethodsDistributed2009, martinsMultidisciplinaryDesignOptimization2013}.

The MDAO architecture can have significant impact on optimisation performance, in terms of achieved optimality, computational efficiency and robustness~\cite{alexandrovInitialResultsMDO1998,yiComparisonMDOMethods2008,tedfordBenchmarkingMultidisciplinaryDesign2010, grayStandardPlatformBenchmarking2013}.
In particular, usage of an (in)appropriate optimisation architecture for the problem at hand can lead to significant differences in computational efficiency, because the architecture determines the structure of iteration loops in the code, which drive computational cost.
As the integration of more disciplines rapidly increases the possibilities for different arrangements, architectural considerations become more important. 
Any such considerations first of all require insight into the applied architectures, further motivating the need for increased transparency of MDAO codes.

To this end, Extended Design Structure Matrix (xDSM) visualisations are commonly applied to provide an overview of the structure of MDAO applications~\cite{lambeExtensionsDesignStructure2012, martinsEngineeringDesignOptimization2021}. 
xDSMs are a specific version of a Design Structure Matrix (DSM), which is used more generally to represent the structure of complex systems~\cite{eppingerDesignStructureMatrix2012}.
For MDAO codes, the overview provided by such visualisations allows insight into the MDAO architecture, facilitating its improvement.
In fusion, the structure of optimisation codes is typically not visualised in a systematic way, and often it is unclear what choices are made in the structure of the code. 
For digital reactor design approaches, this is critical to maintain transparency and efficiency of the used analysis and optimisation codes. 

In this work, we position fusion integrated design approaches as applications of the broader MDAO field, allowing transfer of literature best practices.
We focus in particular on the application of DSM methods, demonstrating how they can be used to coordinate the integration analysis and optimisation codes in fusion.
As we will show with two case studies, DSM methods support analysis of structural assumptions made in model couplings, and of cost-driving feedback loops appearing from those couplings.
Furthermore, we implement a minimally reconfigured code structure demonstrating that architectural considerations based on an automatically generated DSM can reduce model evaluations by 20--50\%.
As we will argue, the impact of such architectural considerations is particularly important for future reactor design frameworks that include expensive higher fidelity models.

To this end, we will first introduce some background in \cref{sec:background}, to highlight the state of the art in the DSM and MDAO fields that motivate this work. 
In \cref{sec:method} we discuss the applied methods, and introduce the two case studies: stellarator optimisation with SIMSOPT~\cite{landremanSIMSOPTFlexibleFramework2021,jorgeSingleStageStellaratorOptimization2023} and the PROCESS systems code~\cite{kovariPROCESSSystemsCode2014,kovariPROCESSSystemsCode2016}, the results of which will be presented in \cref{sec:results}. 
Based on these cases, \cref{sec:discussion} discusses the applicability of MDAO and the value of DSM methods to increase transparency and efficiency of fusion reactor design codes.

\section{Background}
\label{sec:background}
\subsection{Multidisciplinary Design Analysis and Optimisation}
\label{subsec:MDAO}

In both aerospace and fusion engineering design, optimisation algorithms are applied to help find designs that satisfy the stakeholder requirements by searching for a set of design variables which optimises the performance of the system.

Although aerospace and nuclear fusion systems are distinctly different applications, the formulation of the underlying mathematical optimisation problem is application agnostic~\cite{martinsEngineeringDesignOptimization2021}.
Mathematically, all optimisation problems can be formulated as minimising an objective function $f(\mathbf{x})$ with respect to design variables $\underline{\mathbf{x}} \leq \mathbf{x} \leq \overline{\mathbf{x}}$.
The set of acceptable designs can be limited by applying (in)equality constraints $\mathbf{c}(\mathbf{x})$, which are typically used to represent models of the governing physics and engineering limits.

For optimisation of large-scale complex systems, this formulation is not particularly well-suited.
In such cases, the design process is typically a collaborative effort with experts using specific disciplinary analysis models to design specific subsystems.
When these disciplines and subsystems are integrated into a complete system, these models typically need to be coupled to resolve trade-offs occurring both within disciplinary subsystem domains and across their boundaries. 

Optimisation in such environments is in the domain of Multidisciplinary Design Optimisation (MDO)~\cite{sobieszczanski-sobieskiMultidisciplinaryDesignOptimization2015}, which can be  defined as a "methodology for the design of systems in which strong interaction between disciplines motivates designers to simultaneously manipulate variables in several disciplines"~\cite{sobieszczanski-sobieskiMultidisciplinaryAerospaceDesign1997}.
The MDO field has developed methods to enable increased scale and multidisciplinarity of optimisation problems in aerospace, which require advanced methods to address the associated organisational and computational challenges~\cite{sobieszczanski-sobieskiMultidisciplinaryAerospaceDesign1997}.
Similar methods apply to integrated analysis without optimisation as well; therefore we refer to this field more generally as Multidisciplinary Design Analysis and Optimisation (MDAO), following Gray et al.~\cite{grayOpenMDAOOpensourceFramework2019}.

In the MDAO methodology the general optimisation problem formulation is decomposed into $N$ separate disciplines (sometimes referred to as \textit{components}, \textit{models} or \textit{modules}).
As an optimisation problem can be decomposed equally well in terms of physical components or disciplines~\cite{tosseramsClassificationMethodsDistributed2009}, we represent these generally as \textit{analysis models}, reserving the term discipline to refer to a field of expertise.
The decomposed problem formulation and organisational coordination strategy for a specific application is referred to as the MDAO architecture, which essentially defines the coupling between models and their integration in the overarching optimisation~\cite{martinsMultidisciplinaryDesignOptimization2013}.
The software implementation of an MDAO architecture, combining the selected models and the drivers (i.e.\ optimisers and solvers) that coordinate them, is referred to as an MDAO system~\cite{grayOpenMDAOOpensourceFramework2019, ciampaAGILEParadigmNext2020}.

The multidisciplinary optimisation problem can then be generally formulated as the `all-at-once' problem in \cref{eq:mdo_problem_formulation}~\cite{martinsMultidisciplinaryDesignOptimization2013}. 
\begin{equation}
\label{eq:mdo_problem_formulation}
\begin{array}{rl}
\textnormal{minimise} & f_0(\mathbf{x}, \mathbf{y}) + \sum_{i=1}^{N} f_i(\mathbf{x}_0, \mathbf{x}_i, \mathbf{y}_i) \\
\textnormal{by varying} & \mathbf{x}, \hat{\mathbf{y}}, \mathbf{y}, \bar{\mathbf{y}} \\
\textnormal{subject to} & \mathbf{c}_0(\mathbf{x}, \mathbf{y}) \geq 0 \\
 & \mathbf{c}_i(\mathbf{x}_0, \mathbf{x}_i, \mathbf{y}_i) \geq 0 \\
 & \mathbf{c}^{\mathrm{c}}_i = \hat{\mathbf{y}}_i - \mathbf{y}_i = 0 \\
 & \mathbf{R}_i(\mathbf{x}_0, \mathbf{x}_i, \hat{\mathbf{y}}_{j \neq i}, \bar{\mathbf{y}}_i, \mathbf{y}_i) = 0
\end{array}
\end{equation}
In this formulation the design vector $\mathbf{x} = [\mathbf{x}_0, \mathbf{x}_1, \ldots, \mathbf{x}_N]$ is split into shared design variables $\mathbf{x}_0$ and local design variables $\mathbf{x}_i$ that are associated with model $i \in \{1, \ldots, N\}$.
These models can be defined explicitly or implicitly, and are generally represented with the residual functions $\mathbf{R}_i$.
An implicit model may contain a solver to converge the residual by iterating over local state variables $\bar{\mathbf{y}}_i$, or this may be done directly by the optimiser as in \cref{eq:mdo_problem_formulation}.
Coupling variables $\mathbf{y}_i$ represent the output variables of a model that are used by other models $j\neq i$.
Furthermore, shared and local objectives ($f_0(\mathbf{x}, \mathbf{y})$, $f_i(\mathbf{x}_0, \mathbf{x}_i, \mathbf{y}_i)$) and constraints ($\mathbf{c}_0(\mathbf{x}, \mathbf{y})$, $\mathbf{c}_i(\mathbf{x}_0, \mathbf{x}_i, \mathbf{y}_i)$) are distinguished.
In some architectures, variable copies $\hat{\mathbf{y}}_i$ are used that are held to equivalence with consistency constraints $\mathbf{c}^{\mathrm{c}}_i$, allowing models to execute in parallel.
As Martins and Lambe~\cite{martinsMultidisciplinaryDesignOptimization2013} describe, the formulation in \cref{eq:mdo_problem_formulation} is not typically implemented, but forms a basis from which architectures can be derived.
The architecture then defines how the problem is to be solved.

Over the years, various MDAO architectures have been developed, with varying applicability and performance for different types of problems~\cite{martinsMultidisciplinaryDesignOptimization2013, tedfordBenchmarkingMultidisciplinaryDesign2010,tosseramsClassificationMethodsDistributed2009}. 
Monolithic architectures feature a single optimisation algorithm and differ in the way feedback loops are converged.
The Multidisciplinary Feasible (MDF,~\cite{cramerProblemFormulationMultidisciplinary1994}) architecture delegates the couplings between models to a Multidisciplinary Analysis (MDA) solver (which may execute a Gauss-Seidel or Jacobi scheme itself~\cite{martinsEngineeringDesignOptimization2021}), whereas with Individual Discipline Feasible (IDF,~\cite{cramerProblemFormulationMultidisciplinary1994}) the model couplings are handled by the optimiser via consistency constraints. 
In Simultaneous Analysis and Design (SAND,~\cite{haftkaSimultaneousAnalysisDesign1985}), the optimiser also converges the residuals of individual (implicit) models.
Distributed architectures feature multiple nested or alternating optimisers, such as Collaborative Optimisation (CO,~\cite{braunCollaborativeOptimizationArchitecture1996}), Analytical Target Cascading (ATC,~\cite{kimTargetCascadingOptimal2003}), Augmented Lagrangian Coordination (ALC,~\cite{tosseramsAugmentedLagrangianCoordination2008a}) or Asymmetric Subspace Optimisation (ASO,~\cite{chittickAsymmetricSuboptimizationApproach2009}).
Though monolithic architectures appear to be more computationally efficient for most applications~\cite{tedfordBenchmarkingMultidisciplinaryDesign2010,hwangLargescaleMultidisciplinaryDesign2019}, distributed architectures may be particularly valuable if decision authority should remain distributed~\cite{krooMDOLargescaleDesign1997, tosseramsClassificationMethodsDistributed2009, luppMonolithMustWhy2025}.
Design teams can then retain ownership over their optimisation (sub)problem, with an overarching driver algorithm handling the coordination between disciplines.

In the past decades, MDAO has found successful applications in aerospace~\cite{papageorgiouMultidisciplinaryDesignOptimization2018, hwangLargescaleMultidisciplinaryDesign2019}, with for example significant acceleration of the design process demonstrated in an industrial aircraft development case study~\cite{flagerComparisonMultidisciplinaryDesign2007june}.
The development of application-agnostic open source tools, such as OpenMDAO~\cite{grayOpenMDAOOpensourceFramework2019, openmdaoteamOpenMDAO2026} and GEMSEO~\cite{gallardGEMSPythonLibrary2018, gemseoGemseo2025}, now facilitates transfer of state-of-the-art approaches, and MDO approaches have found application in e.g.\ offshore wind~\cite{patryniakMultidisciplinaryDesignAnalysis2022}, automotive~\cite{backrydMultidisciplinaryDesignOptimisation2017} and civil engineering~\cite{muthumanickamMultidisciplinaryDesignOptimization2023} (see also~\cite{hwangLargescaleMultidisciplinaryDesign2019}).

\subsection{Design Structure Matrices}
The Design Structure Matrix (DSM) has a broad history as a visualisation and analysis tool in complex system development~\cite{stewardDesignStructureSystem1981, eppingerDesignStructureMatrix2012}.
Similar to an adjacency matrix~\cite{hararyGraphsMatrices1967}, or N$^2$ chart~\cite{lanoN2Chart1977}, it provides an overview of the interfaces between elements in the system.
DSMs typically enrich this with additional information to yield highly customisable overviews of a system that have been used to represent, for example, products, processes, and organisations~\cite{eppingerDesignStructureMatrix2012}.
DSMs are particularly suited to reveal the structure of complex systems, especially when sequencing (e.g.~\cite{tarjanDepthFirstSearchLinear1972}) or clustering algorithms (e.g.~\cite{wilschutMultilevelFlowBasedMarkov2017}) are used to modify the row/column order of the vertices.
Though DSMs have found initial usage in fusion~\cite{beernaertDiscoveringPotentialDependency2024a}, the authors are unaware of prior DSM-based analyses of integrated reactor design codes.

\begin{figure}[h]
    \centering
    \includegraphics[width=\linewidth]{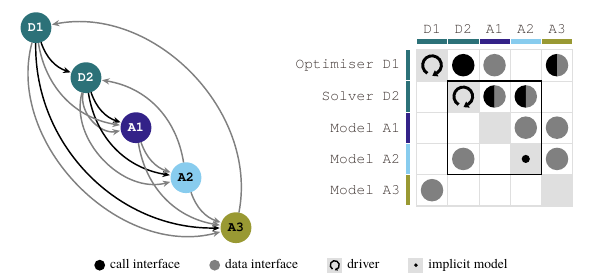}
    \caption{
    An example DSM with a corresponding graph visualisation, representing a Gauss-Seidel MDF architecture.
    Diagonal matrix entries correspond to the vertices of the graph.
    Off-diagonal entries correspond to edges between the vertices.
    The DSMs in this paper follow an input-in-columns convention to represent directed edges, such that for example the entry in the bottom-left corner is a feedback entry from A3 to D1.
    The specific meaning of the different types of vertices and edges in this example is tailored for the PROCESS case study, as detailed in \cref{sec:method}.
    }
    \label{fig:DSM_example}
\end{figure}

As \cref{fig:DSM_example} illustrates, a DSM is essentially a compact visualisation of a graph.
The diagonal entries in the DSM correspond to the vertices of the graph, with the off-diagonal elements representing the edges between them.
Directed edges appear as asymmetric entries in a DSM, where it displays input to vertices in columns and output in rows (i.e. input-in-columns convention).
Correspondingly, all feedforward edges are above the diagonal, and entries below the diagonal may indicate feedback.

In \cref{fig:DSM_example}, we have customised the DSM to distinguish two types of edges between vertices of four different categories.
The different types of vertices have been further differentiated with symbols on the diagonal matrix entries.
A thin black box is added to outline a particular set of vertices, in this case to highlight a specific block containing circular dependencies (\textit{feedback loops}) between vertices.
For large-scale systems, such highlighting can aid interpretation of the structure of the system.

Design structure matrices have been used extensively in the multidisciplinary design optimisation field to display the structure of the problems under consideration.
Building on the work of Steward~\cite{stewardDesignStructureSystem1981}, Rogers~\cite{rogersKnowledgebasedToolMultilevel1989} introduced the DSM as a tool to structure multidisciplinary optimisation problems.
Early examples can be found in e.g. Kroo et al.~\cite{krooMultidisciplinaryOptimizationMethods1994}.
More recently, the Extended Design Structure Matrix (xDSM)~\cite{lambeExtensionsDesignStructure2012} has been proposed as a standardised visualisation approach, which will be discussed in \cref{subsec:xdsm}.
State-of-the-art MDAO frameworks typically allow automatic generation of DSMs, deriving the structure directly from the functional inputs/outputs of models~\cite{m.deborstDecompositionAnalysisMultidisciplinary2016, grayOpenMDAOOpensourceFramework2019, gemseoGemseo2025, gallardGEMSPythonLibrary2018, onodiOnodipOpenMDAOXDSM2024}.

Aside from their communicative purpose, DSMs have been used in conjunction with various graph algorithms to structure the coordination of the disciplinary models \cite{altusGeneticAlgorithmScheduling1996, rogersDeMAIDGAEnhanced1996, luGraphPartitioningBasedCoordination2012, jungDecompositionMethodExploiting2013,m.deborstDecompositionAnalysisMultidisciplinary2016}.
This allows to decompose the problem into smaller subproblems, find optimal execution order, or identify parallelisation opportunities \cite{rogersKnowledgebasedToolMultilevel1989}.

\subsection{The Extended Design Structure Matrix}
\label{subsec:xdsm}
With the wide variety in MDAO architectures and applications, the need has arisen to represent MDAO architectures in a standardised way to facilitate communication between experts and non-experts alike.
To this end, the Extended Design Structure Matrix (xDSM) has been developed, simultaneously showing the modular analysis models, optimisers and solvers, and their data and process flow interdependencies \cite{lambeExtensionsDesignStructure2012}.
In this way it provides a simple, clear and information-dense visualisation which directly reflects the mathematical and computational structure of an MDAO architecture.
As this has become a de facto standard, we will now introduce it in more detail.

\begin{figure}
    \centering
    \includegraphics[scale=0.8]{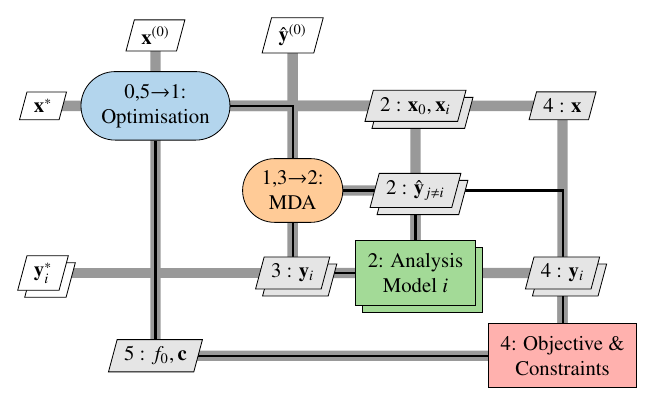}
    \caption{An xDSM visualisation of an MDF optimisation architecture, featuring $N$ analysis models (implicit or explicit) producing coupling variables $\mathbf{y}_i$, iterated to convergence by an MDA solver, and a model computing objective and constraints.
    The MDA solver employs a Jacobi scheme, introducing variable copies $\hat{\mathbf{y}}_i$ such that models can execute in parallel.
    Objectives and constraints are computed after MDA convergence, in an outer loop driven by the optimiser. 
    Note that $^{(0)}$ and $^*$ are used to represent initial and optimal values respectively.}
    \label{fig:xdsm_example}
\end{figure}

As exemplified by the MDF architecture visualised in \cref{fig:xdsm_example}, the xDSM shows computational elements (models and drivers) as the diagonal entries of the matrix, with interfaces as off-diagonal entries.
Analysis models are visualised as rectangles, prompted by a \textit{driver} (rounded rectangle) such as an optimisation algorithm, a design-of-experiments driver to scan over a range of inputs, or an MDA solver to converge a fully coupled set of analysis models.
Note that we consider objectives and constraint computation to be (explicit) analysis models as well, though this is not always done.
Data flow is visualised using the thick grey lines, with off-diagonal entries (referred to as \textit{data interfaces}) revealing the exchanged variables, using the same input-in-columns convention as the regular DSM.
The execution order is shown in the numbering of the entries, visually aided by thin black lines, which we refer to as \textit{process interfaces}.
The iterative nature of drivers is indicated using an arrow; for example, 0,5$\rightarrow$1 indicates that both steps 0 and 5 are followed by step 1 until convergence is reached.
In this way the xDSM shows both data and process interfaces, providing a full overview of the implemented MDAO system.

In comparison with the DSM visualisation in \cref{fig:DSM_example}, which displays a similar architecture, the xDSM shows more detail regarding the mathematical structure of the problem. 
However, because the DSM omits the variable names and moves vertex names to the row labels, it is significantly more compact.
Therefore, we will employ both visualisations in this work, depending on the needs of the application case.

\section{Method}
\label{sec:method}
To demonstrate the added value of these methods from the MDAO and DSM fields for the fusion domain, two case studies have been conducted: stellarator optimisation and the PROCESS systems code.
For each case study an (x)DSM has been constructed so that the MDAO architecture can be analysed.
In this analysis, we identify structural assumptions (as opposed to assumptions within individual models) using the absence of interfaces between models, and reveal feedback loops as circular dependencies.
Based on this analysis, we make recommendations to improve the architecture. 
For the PROCESS case, some of these recommendations have been implemented to measure their effect.
We will now introduce these cases and describe how the (x)DSMs have been constructed.

In stellarator reactor design, optimisation is commonly used to find desirable shapes of the magnetic equilibrium and coils that produce it.
Traditionally, the stellarator design process features a two-stage approach, starting with optimisation of the magnetic equilibrium shape for plasma performance, followed by coil optimisation to produce the desired magnetic field with manufacturable coils.
Though plasma and coil design are typically considered separate disciplines, these problems are often solved in close collaboration to balance plasma physics and coil engineering performance.
Recently, single-stage approaches have been pursued to integrate both optimisation problems~\cite{hennebergCombinedPlasmaCoil2021,jorgeSingleStageStellaratorOptimization2023}.
As such, the design of stellarator configurations provides a clear two-discipline optimisation problem.

In this case study, we assess a stellarator optimisation approach by Jorge et al.~\cite{jorgeSingleStageStellaratorOptimization2023}, implemented in SIMSOPT~\cite{landremanSIMSOPTFlexibleFramework2021}.
We focus in particular on their application optimising quasi-helical stellarators.
To demonstrate the applicability of MDAO, the mathematical description of Jorge et al.\ has been reformulated in MDAO terms.
Based on this description, an xDSM has been created for both aforementioned single-stage and two-stage approaches.
The execution order has been extracted from the specific software implementation of Jorge et al.~\cite{jorgeRogeriojorgeSingle_stage_optimization2024}.
It should be noted that the defined analysis models do not directly correspond with software modules as implemented in the code.
Furthermore, for the purpose of this example solely the optimisation approach is visualised, excluding post-processing and the implemented warm-start initialisation approach.
Based on these xDSMs, we will compare both approaches and suggest recommendations for future developments.

PROCESS~\cite{kovariPROCESSSystemsCode2014, kovariPROCESSSystemsCode2016,morrisPROCESS2026} is a leading systems code that has been used for the design of e.g.\ EU-DEMO~\cite{colemanDefinitionEUDEMODesign2025}, STEP~\cite{muldrewConceptualDesignWorkflow2024} and HELIAS~\cite{warmerSystemCodeAnalysis2016,lionSystemsCodeModels2023}.
PROCESS consists of many low fidelity models, which are individually comparatively cheap in execution time, coupled to a VMCON optimiser~\cite{craneSolutionGeneralNonlinear1980}.
However, these models are connected in a non-trivial manner that has evolved over the years due to the changing use cases, leading to a complex MDA architecture.
This makes it a particularly interesting case to analyse with a DSM. 

To handle the large number of models and variables that are involved, we will use the more compact, custom DSM visualisation as introduced in \cref{fig:DSM_example}.
This DSM is semantically similar to an xDSM, differing mostly in visualisation style.
Here we emphasise implicit models (containing local solvers), with a black dot in the diagonal entry.
Off-diagonal matrix entries still represent the execution order and data interfaces, yet compactly as dots. 
Data interfaces no longer display the name of the exchanged variable, though an interactive format can be exported that shows this information in interactive tooltips.
Semantically, the only difference is that the process interfaces (thin black lines) in the xDSM have been replaced with call interfaces.
A \textit{call interface} directly matches the software implementation, representing a computational element (row) calling another element (column), waiting for the called element to finish before continuing its own execution.
If two elements are called by the same element, the execution order is defined with the diagonal position of the vertices in the DSM.
A call interface thus indicates execution as nested calls of computational elements, whereas a process interface in an xDSM display indicates the execution order as a sequence of steps, irrespective of nesting.
This simplifies the resulting visualisation, though it cannot express parallel execution of models.
Consequently, the depicted execution order in \cref{fig:DSM_example} and \cref{fig:xdsm_example} is equivalent, even though a different MDA scheme is used.
Only the A1$\rightarrow$A2 data interface in \cref{fig:DSM_example} distinguishes it as a Gauss-Seidel scheme, as the interface prevents parallel execution.
Lastly, we reorder the diagonal position of models using a Tarjan-based algorithm to reveal the smallest blocks that contain every feedback loop ('strongly connected components'), with a Markov-based tearing method to reduce feedback entries in each block~\cite{tarjanDepthFirstSearchLinear1972, wilschutRatioCASERaGraph2026}.
As call interfaces reflect the originally implemented execution order, they are omitted when models have been reordered.

Due to the large number of involved models, the PROCESS DSM is not constructed by hand, but using a static code dependency analysis. 
This analysis is implemented in a custom parser, which parses the PROCESS source code using Python's abstract syntax tree~\cite{AstAbstractSyntax}.
To that end, first a call graph is constructed by recording which Python function (or method) is called from which function's body. 
Conditional statements (if/else) typically use configuration variables as arguments of the condition, so they are resolved using the PROCESS input file.
If a conditional statement depends on a runtime variable, or fails to be resolved otherwise, all if/else branches are traced.
Based on a depth-first search of the resulting call graph, the model execution order is reconstructed, using the location of calls (line numbers) to order multiple calls from one function.
Data interfaces are created between individual functions based on the function arguments (input) and return statements (output), as well as local reads/writes to PROCESS' shared data structure.
For cases where the function level is too detailed, functions are aggregated based on the Python class or file they belong to.
In PROCESS, these classes are typically referred to as models; we will refer to the aggregation of functions in a single file as a model as well.
Drivers are declared manually in source code comments, as their implementation is distributed over multiple functions in multiple files.

The parser is constructed using Claude Code Opus and Fable models (Anthropic) in an interactive process that has converged to the presented DSMs by continuous review of code output and construction of an extensive test suite to prevent code changes breaking validated results (i.e.\ regression).
Due to the fundamental limits of static analysis, such as conditional statements dependent on runtime variables, the resulting DSM cannot be guaranteed to be fully accurate.
To further increase confidence in the results, a runtime tracer is implemented in parallel to verify the extracted interfaces during code execution.
Final validation is done manually, using comments written by the parser into the PROCESS source code.
This allows checking directly if variable reads/writes and function calls are interpreted correctly.
Finally, recommendations based on the DSM have always been checked with the source code.

To implement the recommendations derived from the DSM, an experiment harness is constructed, minimally modifying the PROCESS code.
The specific changes made to the PROCESS code will be described in \cref{subsec:surgery}, as they follow from the DSM analysis.
The modified PROCESS code and the code parser can be found in the reproduction package.

\section{Results}
\label{sec:results}

\subsection{Case 1: stellarator optimisation}
\label{subsec:stellopt}
In this section we will analyse both the two-stage and single-stage approaches as carried out by Jorge et al.~\cite{jorgeSingleStageStellaratorOptimization2023}. 
First, we will formulate both problems in terms of \cref{eq:mdo_problem_formulation} to then analyse the implemented problems with xDSMs, leading to recommendations for future developments.

\subsubsection{Optimisation problem definition}
In the first stage, the plasma performance is optimised by varying the shape of the last closed flux surface $S$, which is parameterised with Fourier coefficients $\mathbf{x}_{\mathrm{surf}}$.
Jorge et al.\ use the objective function $f_{\mathrm{p}}$ in \cref{eq:5_qs_objective}, containing a quasisymmetry term $f_{\mathrm{QS}}$ and a target aspect ratio term that is implemented as a quadratic penalty objective $f_A$ with $\omega_A$ as weight\footnote{Though present in the implementation, Jorge et al.\ omit this factor in their publication as it remains at a default value of 1~\cite{jorgeSingleStageStellaratorOptimization2023,jorgeRogeriojorgeSingle_stage_optimization2024}.}.
The quasisymmetry term is a proxy to minimise neoclassical transport losses, developed by Landreman and Paul~\cite{landremanMagneticFieldsPrecise2022}.
To compute this, the magnetic equilibrium $\mathbf{B}$ is solved for the given surface $S(\mathbf{x}_{\mathrm{surf}})$ using VMEC~\cite{hirshmanThreedimensionalFreeBoundary1986}.
The aspect ratio $A$ and local error with respect to perfect quasisymmetry $\delta \text{QS}_j$ are then computed.
A summation of the mean square error over flux surfaces $s_j$ then yields objective $f_{\mathrm{QS}}$.

\begin{align}
    \label{eq:5_qs_objective}
    \begin{split}
    f_{\mathrm{p}}\left(\mathbf{x}_{\mathrm{surf}}\right)=&f_{\mathrm{QS}}\left(\mathbf{B}\right)+\omega_A f_A\left(A\right)\\
    =& \sum_{s_j}\left\langle\left(\delta \text{QS}_j\left(\mathbf{B}\right)\right)^2\right\rangle\ + \omega_A\left(A-A_{\mathrm{target}}\right)^2
    \end{split}
\end{align}

In the second stage, Jorge et al.\ follow the approach of Zhu et al.~\cite{zhuNewMethodDesign2017}.
In this approach the coil shapes are parameterised with Fourier coefficients, which together with the coil currents form the design variables vector $\mathbf{x}_{\mathrm{coils}}$.
The coil objective $f_{\mathrm{c}}$ consists of a quadratic flux term $f_{\mathrm{QF}}$ and a coil engineering term $f_{\mathrm{cc}}$.
The quadratic flux represents the deviation of the produced magnetic field from the desired field optimised in the first stage.
As shown in \cref{eq:5_quadratic_flux}, it is defined as the squared normal magnetic field integrated over the last closed flux surface $S\left(\mathbf{x}_{\mathrm{surf}}\right)$ with unit normal $\mathbf{n}$.
In this way, $f_{\mathrm{QF}}$ penalises inconsistency of the shared variable ($\mathbf{x}_{\mathrm{surf}}$) between the plasma and the coils.
The produced field $\mathbf{B}_{\mathrm{ext}}\left(\mathbf{x}_{\mathrm{coils}}\right)$ is computed on $S$ using the Biot-Savart law.
The coil engineering terms include non-linear expressions to limit coil length ($g_L$), maximum and mean squared coil curvature ($g_{\kappa, \max}$ and $g_{\kappa, \mathrm{msc}}$, respectively), minimum coil separation ($g_d$) and variance in length of sections of an individual coil ($g_{\ell}$), effectively regularising the ill-posedness of the problem.
Though the engineering terms (except $g_{\ell}$) function as design constraints, they are added as penalties with weights $\omega$ to the objective function, as shown in \cref{eq:5_coil_constraints}.

\begin{equation}
\label{eq:coil_obj}
    f_{\mathrm{c}}=f_{\mathrm{QF}} + f_{\mathrm{cc}}
\end{equation}
\begin{equation}
\label{eq:5_quadratic_flux}
f_{\mathrm{QF}}\left(\mathbf{x}_{\mathrm{coils}}, \mathbf{x}_{\mathrm{surf}}\right) =\int_S\left(\frac{\mathbf{B}_{\mathrm{ext}} \cdot \mathbf{n}}{\left|\mathbf{B}_{\mathrm{ext}}\right|}\right)^2 d S
\end{equation}
\begin{align}
    \label{eq:5_coil_constraints}
    \begin{split}
    f_{\mathrm{cc}}\left(\mathbf{x}_{\mathrm{coils}}\right)&=\omega_L g_L+\omega_{\kappa, \max } g_{\kappa, \max}\\
    &+\omega_{\kappa, \mathrm{msc}} g_{\kappa, \mathrm{msc}}+\omega_d g_d+\omega_{\ell} g_{\ell}
    \end{split}
\end{align}

Jorge et al.\ have developed an optimisation approach based on fixed-boundary equilibria, which integrates both optimisation stages into a so-called single-stage optimisation.
As shown in \cref{eq:5_single_stage}, the objective functions are combined with a weight $\omega_{\mathrm{c}}$ given to the coil objective.
Furthermore, the design vector $\mathbf{x}=[\mathbf{x}_{\mathrm{surf}},\mathbf{x}_{\mathrm{coils}}]$ now contains the degrees of freedom of both the plasma and coil design.
Though the models are exactly the same, they show that improved performance could be reached using the single-stage optimisation~\cite{jorgeSingleStageStellaratorOptimization2023}. 

\begin{equation}
    \label{eq:5_single_stage}
f\left(\mathbf{x}_{\mathrm{coils}}, \mathbf{x}_{\mathrm{surf}}\right)=f_{\mathrm{p}}+\omega_{\mathrm{c}} f_{\mathrm{c}}
\end{equation}

With the unification of the design vector and objective, Jorge et al.\ have essentially combined two originally separate optimisation problems into one problem that can be expressed as a multidisciplinary problem following \cref{eq:mdo_problem_formulation}.
The step to an MDAO interpretation requires a strict separation of individual models with explicit inputs and outputs, which can be done in multiple ways. 
For the DSM analysis below we have chosen to decompose the coil and plasma disciplines further, defining VMEC, a Biot-Savart model and the functions computing the four individual objective terms in \cref{eq:5_qs_objective,eq:coil_obj} as $N=6$ individual models.
In this view, we can consider $\mathbf{x}_{\mathrm{surf}}$ and $\mathbf{x}_{\mathrm{coils}}$ as shared design variables $\mathbf{x}_0$, and $A$, $\mathbf{B}$ and $\mathbf{B}_{\mathrm{ext}}$ as a set of coupling variables $\mathbf{y}$ computed by VMEC (implicitly) and a Biot-Savart model (explicitly).
As noted above, $f_A$ and $f_{\mathrm{cc}}$ may be interpreted as design constraints $\mathbf{c}(\mathbf{x},\mathbf{y})$ that have been implemented using a penalty method.

\subsubsection{Analysis using xDSM visualisation}

\begin{figure*}[!t]
\centering
\subfloat[Two-stage stellarator optimisation.]{\includegraphics[scale=0.72]{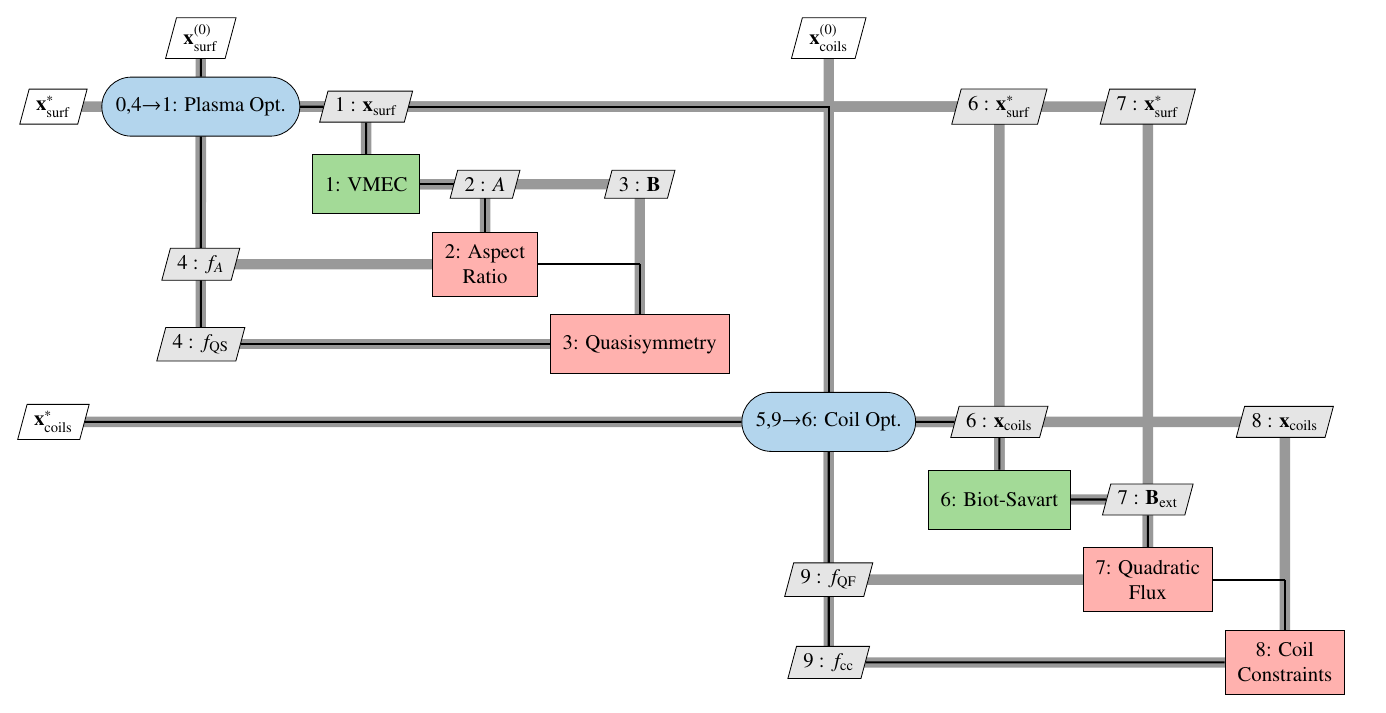}%
\label{fig:two_stage}}
\hfil
\subfloat[Single-stage stellarator optimisation.]{\includegraphics[scale=0.72]{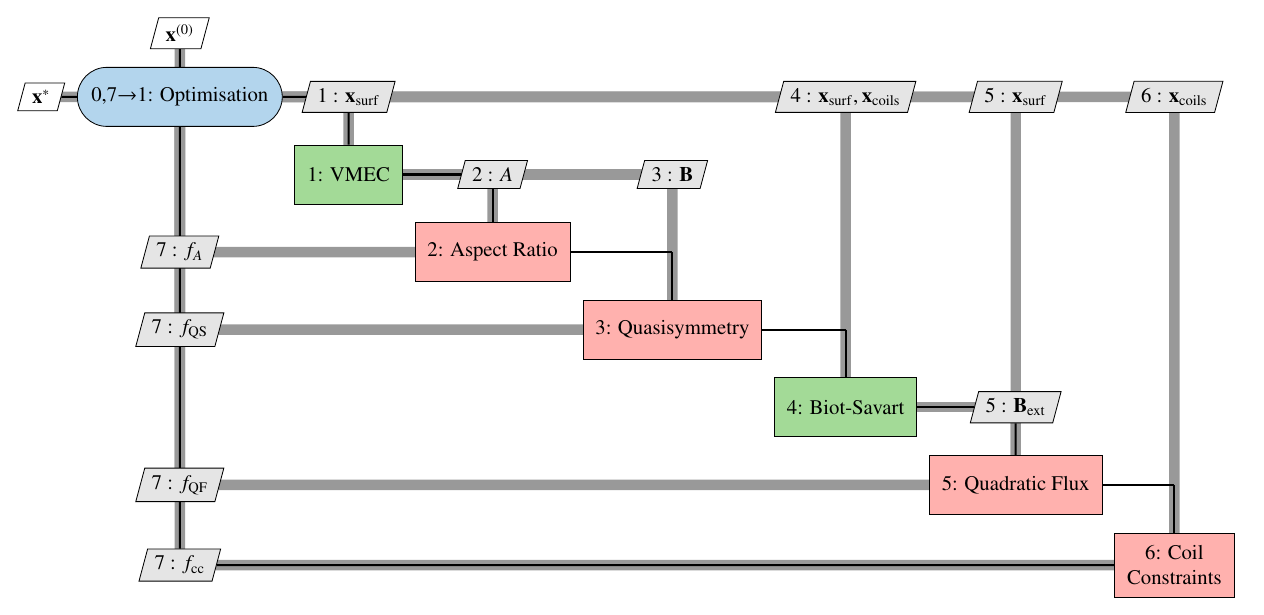}%
\label{fig:single_stage}}
\caption{Extended design structure matrix visualisations of the two stellarator optimisation approaches as implemented by Jorge et al.~\cite{jorgeSingleStageStellaratorOptimization2023}.}
\label{fig:stellarator_xdsm}
\end{figure*}

Let us now turn to the xDSM visualisations of this optimisation problem.
The classical two-stage approach is shown in \cref{fig:two_stage}, with in the top left the first-stage plasma optimisation, which provides the optimal surface $\mathbf{x}_{\mathrm{surf}}^*$ to the second-stage coil optimisation in the bottom right.
This clearly reflects the sequential nature of the two-stage approach: the only dependency is the optimal plasma shape $\mathbf{x}^*_{\mathrm{surf}}$, which defines the surface on which the produced magnetic field $\mathbf{B}_{\mathrm{ext}}$ and corresponding quadratic flux $f_{\mathrm{QF}}$ are computed.

\Cref{fig:two_stage} reveals the assumptions concerning the implemented interface.
Firstly, the lack of dependencies in the bottom left of \cref{fig:two_stage} indicates that the plasma optimisation is independent of the coil optimisation.
Furthermore, it reveals that the quasisymmetry objective $f_{\mathrm{QS}}$ is not computed based on the magnetic field produced by the coils, but based on the desired magnetic surface $\mathbf{x}_{\mathrm{surf}}$.
Any inaccuracies in the production of this surface, as computed by $f_{\mathrm{QF}}$, are thus assumed negligible.
This is a significant issue in stellarator optimisation, as it is not guaranteed that sufficiently performing coils can be designed for a desired magnetic surface, risking expensive reiteration of the plasma optimisation~\cite{jorgeSingleStageStellaratorOptimization2023}.

The single-stage approach addresses the neglected dependency of plasma performance on coil optimisation by integrating both optimisation stages into a single optimisation, as shown in  \cref{fig:single_stage}.
The quadratic flux objective $f_{\mathrm{QF}}$ can now influence quasisymmetry $f_{\mathrm{QS}}$ by changing $\mathbf{x}_{\mathrm{surf}}$ in the next iteration, changing the last closed flux surface such that it can be produced more easily.
The optimiser balances the surface shape according to both plasma and coil performance, a trade-off that is governed by weight $\omega_{\mathrm{c}}$ in the combined objective (\cref{eq:5_single_stage}).
Note that the quasisymmetry of the optimal design point still does not contain a correction due to the imperfect coil design.
As \cref{fig:single_stage} shows, the quasisymmetry $f_{\mathrm{QS}}$ is still computed with the idealised surface $\mathbf{x}_{\mathrm{surf}}$, without a dependency on $\mathbf{B}_{\mathrm{ext}}$ or $f_{\mathrm{QF}}$.
The final results of the optimisation thus still depend on the assumption that the coil optimisation perfectly reproduces the magnetic field.
However, the optimiser is now able to reduce this error by changing the surface shape, at the cost of the other objective terms.
The assumption does not hold exactly if $f_{\mathrm{QF}}$ receives a finite weight; therefore Jorge et al.~\cite{jorgeSingleStageStellaratorOptimization2023} check the discrepancy with a post-processing step.

Aside from comparing both stellarator optimisation approaches, \cref{fig:stellarator_xdsm} enables the approaches to be evaluated in terms of the broader MDAO literature.
For example, \cref{fig:two_stage} clearly shows the two optimisation stages by which it can be classified as a sequential approach with $\mathbf{x}_{\mathrm{surf}}$ connecting both problems.
Such a sequential approach tends to yield suboptimal design points~\cite{sobieszczanski-sobieskiMultidisciplinaryAerospaceDesign1997,martinsEngineeringDesignOptimization2021}, though it is typically less expensive computationally.
The single-stage approach can then essentially be considered a monolithic MDF architecture with a trivial, single-pass MDA, as all of the model couplings are feedforward.
The coupling between the models is implemented via the shared design variables and resolved by balancing the competing objectives.
The results of Jorge et al.~\cite{jorgeSingleStageStellaratorOptimization2023} then confirm that a monolithic architecture can improve results, though at higher computational cost.
For a specific problem, the value of a monolithic architecture then depends on the required performance, and the impact of the assumption that would be needed to decouple the disciplines to allow a sequential approach.

For optimisation problems where models are directly interdependent and couplings have to be resolved fully consistently, the distinctions between different MDAO architectures become relevant.
Considering the high complexity and performance requirements of fusion reactors, many cases may be considered where a decoupling such as in \cref{fig:two_stage} is no longer adequate.
For example, consider the inclusion of finite-beta plasma scenarios requiring the magnetic equilibrium to be solved self-consistently with profiles computed by a transport model.

As introduced in \cref{subsec:MDAO}, an MDF architecture addresses this coupling with an MDA solver iterating over the feedback loop, whereas an IDF architecture cuts the feedback loop and enforces consistency with a constraint.
In other cases, it may be preferable to avoid nested solvers to improve computational efficiency.
In such cases the solvers of implicit models (such as VMEC in \cref{fig:stellarator_xdsm}) can be absorbed in the optimiser (i.e.\ a SAND architecture).
This can be more efficient, but requires access to the model residuals, which may involve significant development effort~\cite{tedfordBenchmarkingMultidisciplinaryDesign2010,martinsMultidisciplinaryDesignOptimization2013}.
In stellarator optimisation, such an architecture can be recognised in the DESC-based approach of Conlin et al.~\cite{conlinStellaratorOptimizationConstraints2024}.
Lastly, distributed architectures can be considered if it is a priority that design authority remains with disciplinary design teams.
For example, the second-stage coil optimisation could be executed in every plasma optimisation iteration, to minimise the relatively expensive VMEC evaluations.
This would resemble an ASO architecture~\cite{chittickAsymmetricSuboptimizationApproach2009}.

With the analysis of two common stellarator optimisation approaches, we show that these problems are indeed applications of the broader MDAO field. 
This allows literature-based suggestions for alternative architectures to resolve couplings between the models consistently.
In particular for stellarator optimisation, it may help clarify the coordination of plasma and coil optimisation, and an extensive mapping of MDAO concepts to e.g.\ the work of Henneberg et al.~\cite{hennebergCombinedPlasmaCoil2021} may yield valuable insights.
The usage of the xDSM in this analysis illustrates its value in providing an overview of an implemented approach, directly revealing the differences between architectures.
Furthermore, the absence of matrix entries can reveal assumptions made in the couplings between models.
Such overviews become more valuable when applications include more models, as we will now illustrate with the larger-scale PROCESS case study.

\begin{figure*}[!t]
    \centering
    \makebox[\textwidth][r]{\includegraphics[width=1.06\textwidth]{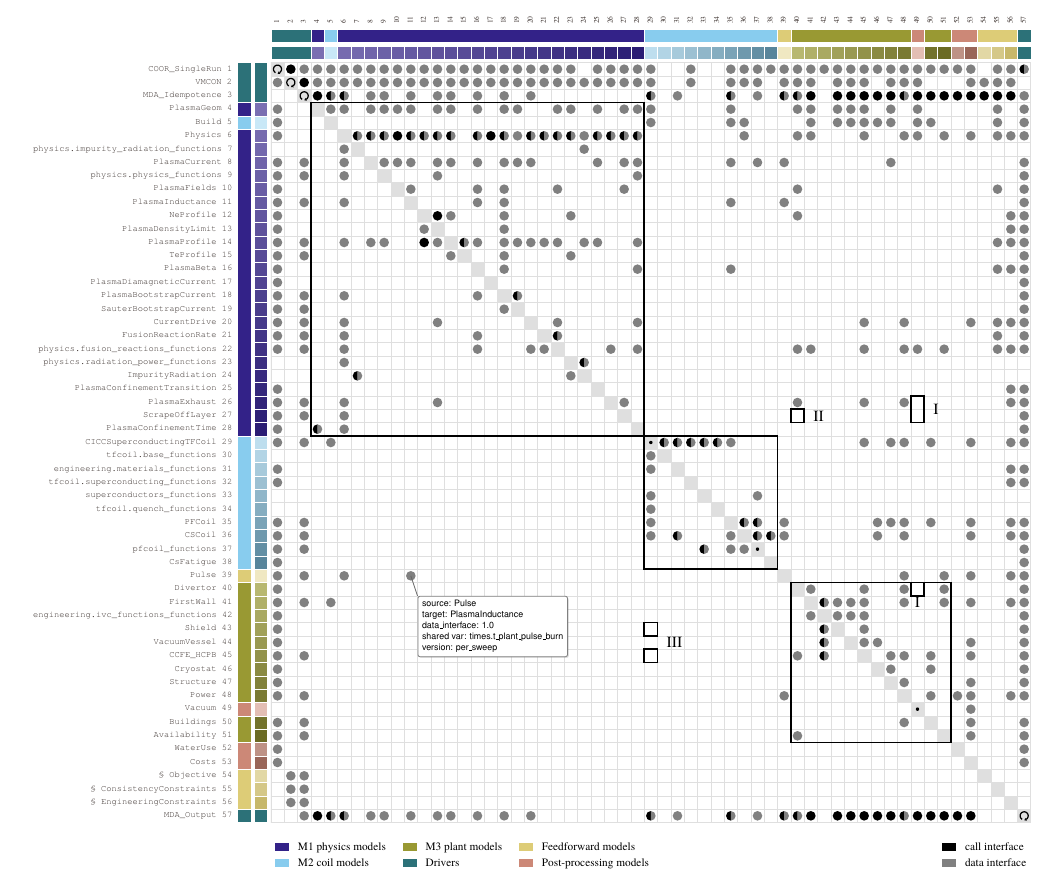}}%
    \caption{A DSM visualisation of the MDAO architecture of the PROCESS code for a large tokamak configuration, as extracted from the source code using a static code dependency analysis.
    Diagonal entries represent the drivers and the PROCESS models, in order of first execution.
    Off-diagonal entries contain data interfaces between models, and call interfaces that represent the execution order.
    Three black boxes are added to highlight the structure of the model couplings as three main modules, and individual cells (\MakeUppercase{\romannumeral 1-\romannumeral 3}) are highlighted where the lack of an interface shows an assumption.
    The colour coding shows how models are categorised for the architectural change in \cref{subsec:surgery}, which reorders them to allow the three modules to be converged separately.
    }
    \label{fig:tokamak_collapsed}
\end{figure*}

\subsection{Case 2: PROCESS}
\label{subsec:process}
For the second case study, the PROCESS code is analysed to create a DSM overview of the implemented architecture.
Firstly we analyse the DSM, resulting in clear suggestions to improve the architecture.
These suggestions are subsequently implemented to demonstrate the potential impact of architecture on code performance.
Lastly, this case study will analyse the architecture at the level of individual Python functions  to assess the potential impact of a more thorough refactoring of the PROCESS MDAO architecture.

\subsubsection{DSM analysis}
\Cref{fig:tokamak_collapsed} displays a DSM generated from a static code dependency analysis of PROCESS.
For the analyses, the figure has been rendered interactively, as exemplified with the \texttt{Pulse} to \texttt{PlasmaInductance} data interface hover tooltip.
The DSM has been modified further to highlight some initial analysis results.
Drivers and models have been colour coded on the DSM axis, three tightly coupled modules (M1, M2, M3) have been outlined with large black boxes, and small boxes (I-III) highlight missing interfaces indicating an assumption.
We will now discuss these insights in further detail.

Firstly, we consider the individual interfaces shown in the DSM.
As in the previous case study, the (absence of) data interfaces between models can reveal assumptions made in the code.
Three assumptions have been highlighted to illustrate how the DSM in \cref{fig:tokamak_collapsed} can be used to find assumptions underlying the results:
\begin{enumerate}[label=\Roman*,leftmargin=*]
\item It is noted that the \texttt{Vacuum} model does not take any input from the \texttt{PlasmaExhaust}, \texttt{ScrapeOffLayer} or \texttt{Divertor} models.
Required vacuum pump speeds are computed indirectly, based on geometry terms (\texttt{Build}, \texttt{PlasmaGeom}), total fusion power (\texttt{physics.fusion\_reactions\_functions}) and required fuelling rate (\texttt{Physics}).
From a design perspective, this means exhausted impurities from \texttt{FirstWall} and \texttt{Divertor} (perhaps via \texttt{ScrapeOffLayer}) and neutral compression in the \texttt{Divertor} have not been taken into account.
In fact, closer inspection reveals that no particle exhaust model is implemented at all, with divertor chamber pressure being a user input.
\item From the DSM, it appears the \texttt{ScrapeOffLayer} does not pass data to the \texttt{Divertor} model, whereas it would be expected that the Scrape-Off Layer (SOL) width (or heat flux derived from it) would be used by the \texttt{Divertor}. 
The cause appears to be that both the \texttt{ScrapeOffLayer} and the \texttt{Divertor} compute the SOL width, with both values used for different purposes.
\item In a third case, we note that the blanket models \texttt{CCFE\_HCPB} and \texttt{Shield} do not pass data to the \texttt{CICCSuperconductingTFCoil} model (or any other coil model). 
The nuclear heating in the toroidal field coils is an output of \texttt{CCFE\_HCPB}, but is not passed to the coil model for further thermal computations.
Furthermore, a designer may expect a blanket model to compute the peak fast neutron fluence passing through the blanket to the coils. 
Instead, the \texttt{CICCSuperconductingTFCoil} uses a user-specified fast neutron fluence limit, so that end-of-life resistivity is computed for a worst case. 
\end{enumerate}

Based on the DSM the optimisation architecture can also be evaluated from the MDAO perspective. 
The DSM shows a structure with one top-level optimiser (\texttt{VMCON}) driving an MDA solver (\texttt{MDA\_Idempotence}) that converges couplings between models, resembling an MDF architecture.
The PROCESS architecture differs from a classical MDF architecture, as some coupling variables are handled by the optimiser as consistency constraints instead of by the MDA (IDF-like).
Furthermore, it features a second post-optimality MDA solver (\texttt{MDA\_Output}), which in principle should be redundant if the optimiser converges properly.
However, as \texttt{MDA\_Idempotence} only converges objectives and constraints, instead of all coupling variables, it is justified as a final, more thorough convergence check.

Additionally, the DSM shows an overview of the MDA architecture as well, revealing the couplings between models.
From the call interfaces a complicated structure of nested model calls appears. 
The data interfaces between the models reveal that the MDA architecture is a Gauss-Seidel scheme: models use variables updated earlier in the same iteration, so cannot be executed in parallel.
Regarding the structure of these interfaces, the DSM reveals three modules consisting of tightly coupled models.
These modules can be identified as physics models (M1), coil models (M2) and plant models (M3), which are each highlighted with a colour group.
Within each module, many circular dependencies can be found.
Between these modules there is only very limited data flowing backwards to models that were executed earlier, suggesting the modules may be easily partitioned.

Note that not all of the circular dependencies in the DSM are actual feedback loops in the code.
Because some of the PROCESS models are composed of functions that are independent, sequential operation of functions distributed over multiple models can appear as a circular dependency in the matrix.
For this reason we will also discuss a DSM containing the individual functions in \cref{subsec:expanded_dsm}.
Regardless, the feedback loops in \cref{fig:tokamak_collapsed} provide an upper bound on the circular dependencies inside and between modules.

Based on \cref{fig:tokamak_collapsed}, specific changes can be suggested to the MDAO architecture of PROCESS. 
In particular, it seems to be advantageous to partition the MDA into the three modules.
Currently, if one specific feedback loop has not yet converged, all other models also need reiteration.
Partitioning can reduce this unnecessary evaluation of models, as each module can be converged independently.
The DSM is now inspected to reveal how such a partitioning could be achieved.

Firstly, as no M1 model takes input from the \texttt{Build} model (\#5), it can be run as part of M2 after M1 has finished.
This sequencing operation confines the feedback entry from \texttt{CICCSuperconductingTFCoil} (\#29) to \texttt{Build} in M2.
This leaves the \texttt{FirstWall} (M3, \#41) to \texttt{Build} (M2) feedback entry, which turns out to be removable by sequencing as well.\footnote{On inspection of the source code, this feedback entry originates from a single function inside \texttt{FirstWall} that is only dependent on user-specified initial values.
As such, the function can be extracted and run before M2, removing the apparent feedback.}

The last remaining feedback entries are then related to the \texttt{Pulse} model (\#39), which produces the plant pulse burn time variable responsible for the data interfaces to \texttt{PlasmaInductance} (M1, \#11), \texttt{Physics} (M1, \#6), \texttt{Power} (M3, \#48) and \texttt{Availability} (M3, \#51).
The coupling via this variable can be cut and delegated to the optimiser, adding a consistency constraint as in an IDF architecture.
The optimiser then uses a duplicate burn time as an additional design variable (an iteration variable in PROCESS terms), to be kept consistent with the value computed by \texttt{Pulse}.
\texttt{Pulse} can then be sequenced after M3 to remove the last feedback entry from \texttt{Power} to \texttt{Pulse}.

The MDA can thus be partitioned into the three modules in \cref{fig:tokamak_collapsed} by reordering the execution of a few functions, and delegating the burn time coupling to the optimiser.
As a collateral benefit, this also results in some models not being part of any MDA iteration ('Feedforward models'). 
Furthermore, some models provide no output to the optimiser, so they can be run once after optimisation ('Post-processing models').
These recommendations require only minor changes to PROCESS, which will be implemented in the following section.

\subsubsection{Impact of minimal architectural changes}
\label{subsec:surgery}

To measure the impact of the suggested changes, PROCESS has been run both in evaluation (i.e.\ MDA only) and optimisation mode with the modified MDA architecture.
In evaluation mode the burn time coupling is kept fixed at a perturbed initial value, to see the impact on model evaluations of only the partitioning of the MDA.
As detailed in \cref{app:surgery_eval}, this yields a reduction in model evaluations of 20--50\% for M1 and M3.
M2 does not use fewer iterations in the partitioned case, so it contains the bottlenecking feedback loops in the unified architecture.
We now discuss in more detail the impact of the MDA change on optimisation mode, which may perform differently due to the additional consistency constraint required to converge the burn time coupling.

For the purpose of the experiment, we consider a ladder of four cases in which adjacent cases differ by separate architectural steps.
Firstly, we define a reference case BR that is unmodified PROCESS in optimisation mode, and a control case B0 that implements a baseline MDA algorithm that is applicable to the partitioned architecture as well.
Because converging on just the objective and constraints is impossible for the partitioned case, B0 modifies the MDA convergence criterion to a relative tolerance of $10^{-6}$ on all model outputs.
The same criterion is used for \texttt{MDA\_Idempotence} and for each of the partitioned MDAs, so that all cases are compared at the same accuracy.
Subsequently case B1 takes the burn time coupling out of the MDA (B1 also removes \texttt{MDA\_Output}, which is redundant once all outputs are converged). 
B2 then adds the sequencing and partitioning of the MDA into three separate MDAs for each module in \cref{fig:tokamak_collapsed}.
The step from B0 to B2 thus gives the total effect of the modified architecture on the optimisation.
An overview of the experiment cases is shown in \cref{tab:switch_matrix_opt}.

\begin{table}[!t]
    \centering
    \footnotesize
    \setlength{\tabcolsep}{3pt}
    \begin{tabular}{l|cccc}
        \hline
         & BR & B0 & B1 & B2 \\
        \hline
        MDA architecture      & unified & unified & unified & partitioned \\
        MDAs converged on & obj.\,\&\,constr. & all outputs & all outputs & all outputs \\
        \texttt{MDA\_Output}            & kept & kept & removed & removed \\
        Burn time coupling & MDA & MDA & optimiser & optimiser \\
        Models sequenced             & no & no & no & yes \\
        \hline
    \end{tabular}
    \caption{Overview of the modifications to PROCESS (optimisation mode) implemented in each experiment case.}
    \label{tab:switch_matrix_opt}
\end{table}

\begin{table*}[!t]
    \centering
    \footnotesize
    \setlength{\tabcolsep}{2pt}
    \begin{tabular}{l|l|c|l|c|c|l}
    \hline
    Configuration & Input file (\texttt{.IN.DAT}) & Models & Objective & Design var. & Constraints & Cross-module coupling \\
    \hline
    Large tokamak (\texttt{tok}) & \texttt{large\_tokamak\_nof} & 52 & min. major radius (\texttt{rmajor}) & 20 $\rightarrow$ 21 & 26 $\rightarrow$ 27 & \texttt{t\_plant\_pulse\_burn} \\
    Low aspect ratio DEMO (\texttt{lad}) & \texttt{low\_aspect\_ratio\_DEMO} & 52 & max. burn time (\texttt{t\_plant\_pulse\_burn}) & 19 $\rightarrow$ 20 & 25 $\rightarrow$ 26 & \texttt{t\_plant\_pulse\_burn} \\
    Spherical tokamak (\texttt{st}) & \texttt{st\_regression} & 52 & max. fusion gain (\texttt{big\_q\_plasma}) & 14 & 18 & none (steady state) \\
    \hline
    \end{tabular}
    \caption{Overview of the three baseline PROCESS configurations used in the experiment.
    The cut of the burn time coupling in the experiment adds a design variable and a constraint for the \texttt{tok} and \texttt{lad} cases.}
    \label{tab:surgery_configs}
\end{table*}

Each experiment case is run 25 times with perturbed initial values for three different machine configurations, a large tokamak (\texttt{tok}), a low aspect ratio DEMO (\texttt{lad}) and a spherical tokamak (\texttt{st}), as characterised in \cref{tab:surgery_configs}. 
The configurations differ in objectives, constraints, design variables and slightly in the set of executed models.
The same MDA partitioning can be applied nonetheless, though the burn time variable has a different role in each machine: for the low aspect ratio DEMO it is also the objective, and the spherical tokamak configuration is steady state so there is no burn time coupling in the first place.

The effect of the architectural change on PROCESS optimisation runs is then shown in \cref{tab:phaseB_results}, displaying the total number of model evaluations of a module over the full optimisation.
Perturbed starting points where any of the experiment cases do not converge have been rejected.
The control case B0 matches the cost of the reference run BR reasonably well, with the exception of \texttt{st} which can be explained by a difference in optimiser iterations (see \cref{tab:phaseB_iterations}).
The burn time cut in B1 does not materially increase the number of model evaluations; in fact, for \texttt{lad} the optimiser seems to converge faster.
Similar to the results for evaluation mode, the modified architecture (B2) then leads to a reduction of model evaluations of 20--50\% over B0.
For \texttt{lad} this can be attributed partly to a reduction in optimiser iterations, which are in general quite sensitive to problem configuration, initialisation and convergence settings.
Interestingly, M2 also shows a reduction of 10--40\% in evaluations that did not surface for evaluation mode, and cannot be explained by a change in optimiser iterations alone.
This is likely because about 95\% of the MDA calls in optimisation mode are finite difference evaluations, each of which perturbs a single design variable (thus affecting only a part of the models), whereas for the evaluation mode experiment all coupling variables are perturbed at once.
Therefore, M2 is no longer the slowest loop for many perturbations, so over a full optimisation it is iterated fewer times as well.

\begin{table}[!t]
    \centering
    \footnotesize
    \setlength{\tabcolsep}{3pt}
    \makebox[\columnwidth][c]{%
    \begin{tabular}{l|cccc|cc}
    \hline
    Module & BR & B0 & B1 & B2 & B2/B0 & B2/B0 med [min, max] \\
    \hline
    \multicolumn{7}{l}{\texttt{tok} ($n = 22$)} \\
    \hline
    M1              & 1977 & 2027 & 2040 & 1388 & 0.68 & 0.69 [0.60, 0.80] \\
    M2              & 1977 & 2027 & 2040 & 1761 & 0.87 & 0.88 [0.76, 1.01] \\
    M3              & 1977 & 2027 & 2040 & 1541 & 0.76 & 0.77 [0.66, 0.89] \\
    Feedforward     & 1977 & 2027 & 2040 & 640 & 0.32 & 0.32 [0.28, 0.37] \\
    Post-processing & 1977 & 2027 & 2040 & 1 & 0.00 & 0.00 [0.00, 0.00] \\
    \hline
    \multicolumn{7}{l}{\texttt{lad} ($n = 11$)} \\
    \hline
    M1              & 8095 & 7859 & 5436 & 3670 & 0.47 & 0.54 [0.08, 4.13] \\
    M2              & 8095 & 7859 & 5436 & 4660 & 0.59 & 0.69 [0.11, 5.24] \\
    M3              & 8095 & 7859 & 5436 & 4273 & 0.54 & 0.63 [0.10, 4.80] \\
    Feedforward     & 8095 & 7859 & 5436 & 1714 & 0.22 & 0.25 [0.04, 1.93] \\
    Post-processing & 8095 & 7859 & 5436 & 1 & 0.00 & 0.00 [0.00, 0.00] \\
    \hline
    \multicolumn{7}{l}{\texttt{st} ($n = 22$)} \\
    \hline
    M1              & 6043 & 5050 & -- & 3250 & 0.64 & 0.72 [0.16, 0.89] \\
    M2              & 6043 & 5050 & -- & 3373 & 0.67 & 0.72 [0.17, 0.96] \\
    M3              & 6043 & 5050 & -- & 3324 & 0.66 & 0.73 [0.17, 0.93] \\
    Feedforward     & -- & -- & -- & -- & -- & -- \\
    Post-processing & 6043 & 5050 & -- & 1 & 0.00 & 0.00 [0.00, 0.00] \\
    \hline
    \end{tabular}}
    \caption{Average number of model evaluations over a full optimisation for the experimental cases as specified in \cref{tab:switch_matrix_opt}, for each configuration in \cref{tab:surgery_configs}.
    The effect of the architecture change is reflected in the B2/B0 ratio of the averages, and the median and [min,max] of the per-run ratios over the $n$ perturbed starts.}
    \label{tab:phaseB_results}
\end{table}

Though these results indicate a reduction in the number of model evaluations, this does not directly translate to an equivalent speedup of runtime.
The actual runtime speedup depends not only on the model evaluations but also on their overhead.
For PROCESS, the individual models are very light (around 0.1--3\,ms per call), and overhead is a significant part of the runtime.
A large part of the overhead is introduced with control case B0, as the convergence on all model outputs takes similar time as the model evaluations themselves (see \cref{app:wallclock time}), mostly cancelling the speedup due to the architecture change.
This overhead can be reduced by implementing a different convergence criterion, but the PROCESS optimiser is quite sensitive to this, which can result in slower convergence and/or termination at a different local minimum. 
This would confound the architectural comparison, but does not affect regular applications, where overhead can be reduced.
The impact of this specific architectural change may not be significant for PROCESS directly, and may be limited to specific configurations.
Nonetheless, the results do demonstrate that architectural considerations can significantly affect the number of model evaluations.

\subsubsection{Potential of complete architecture refactoring}
\label{subsec:expanded_dsm}
\begin{figure*}[!t]
    \centering
    \includegraphics{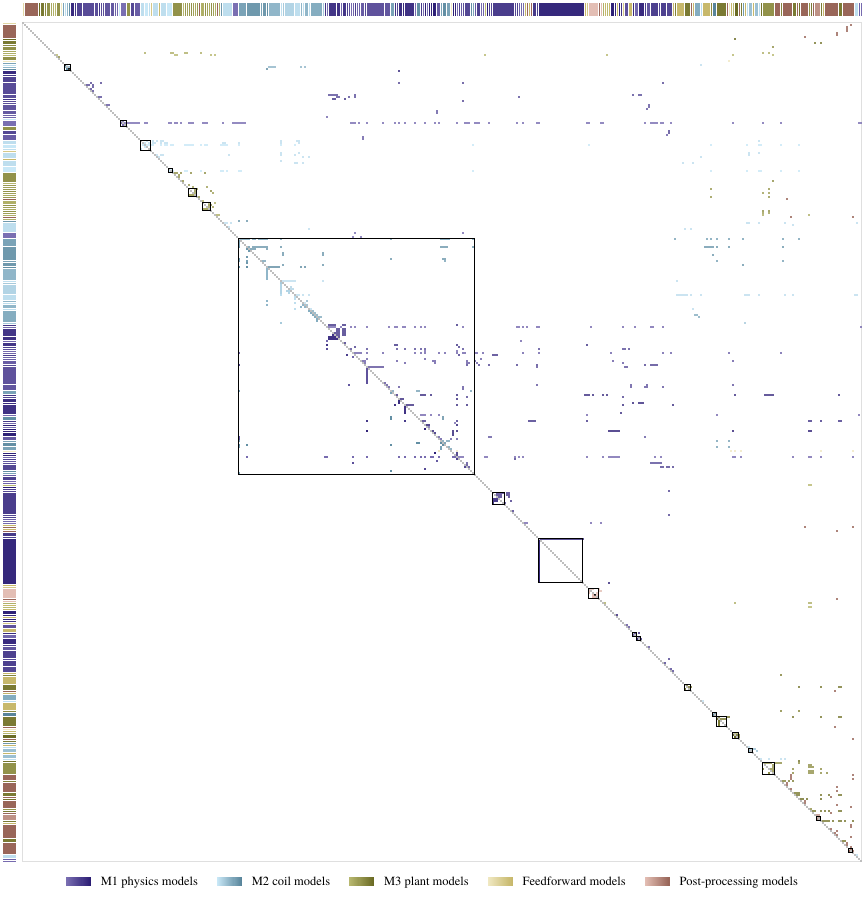}%
    \caption{An expanded DSM overview of the PROCESS code for a large tokamak configuration, showing only data interfaces between individual model functions. 
    This DSM is essentially an expansion of \cref{fig:tokamak_collapsed}, with the colour coding on the axis relating each function to the parent model in \cref{fig:tokamak_collapsed}.
    The functions have been sequenced using a Tarjan-based algorithm~\cite{tarjanDepthFirstSearchLinear1972, wilschutRatioCASERaGraph2026}, revealing that only about half of the PROCESS model functions need to be part of an iterated solver.
    }
    \label{fig:tokamak_expanded}
\end{figure*}

To see what a more rigorous code refactoring may achieve, we now expand the PROCESS DSM, revealing the interconnections between the individual functions that make up the models in \cref{fig:tokamak_collapsed}.
\Cref{fig:tokamak_expanded} shows the expanded DSM of PROCESS without the top-level drivers, and with the functions reordered into `strongly connected components', which are highlighted with black square boxes.
The colour bars on the axis of the DSM then reveal the origin of the individual functions, corresponding to the colours in \cref{fig:tokamak_collapsed}.

As \cref{fig:tokamak_expanded} reveals, roughly half of all functions in PROCESS may be executed sequentially if PROCESS were refactored completely.%
However, the sequenced order is for a specific configuration and may not be advantageous for other configurations.
The limited accuracy of the static code analysis should also be taken into account, so any inferred ordering cannot be implemented directly.
For other codes that have strictly functional models, a direct implementation of the reordered DSM would be possible.
Regardless, \cref{fig:tokamak_expanded} does indicate that a rigorous architectural overhaul of PROCESS may be worthwhile.

In this case study we have applied design structure matrices to analyse the architecture of the PROCESS code and we have implemented some of the recommendations that followed from it.
With the collapsed DSM in \cref{fig:tokamak_collapsed} we have shown that structural assumptions regarding the implemented models can be found in the DSM.
Furthermore, it revealed that it is possible to partition the MDA solver by cutting the burn time coupling, delegating it to the optimiser, and slightly reordering some PROCESS models.
An implementation of these minor changes resulted in a 20--50\% reduction of model evaluations.
As the expanded DSM in \cref{fig:tokamak_expanded} has revealed, this reduction could potentially be increased further with a rigorous code refactoring, as roughly half of the functions may be executed without iteration at all.

\section{Discussion}
\label{sec:discussion}

Future digital design frameworks in fusion are bound to express the complexity of a reactor as intricate couplings between the models.
Motivated by advances in other fields, this work has aimed to introduce methods to deal with the complex couplings in such frameworks.

To that end, integrated design analysis and optimisation approaches in fusion have been reframed as applications of MDAO.
The case studies in \cref{sec:results} show that both systems codes and stellarator shape optimisation used in fusion can be expressed as MDAO problems.
The only thing an MDAO interpretation requires is a strict definition of the boundaries of the models, and the interfaces between them.
Of course, how the models and interfaces are defined remains a design choice.
The defining characteristic of MDAO is then the focus on the couplings between the models, and in particular how feedback loops are resolved with solvers and optimisers.

As we have discussed, this structure of models, solvers and optimisers --- the MDAO architecture --- is typically visualised with an (x)DSM.
The creation of a DSM forces the interfaces between models to be strictly defined,\footnote{If this has not already been done in the development of the software implementation; if it has, the DSM should follow trivially.} and then reveals the resulting structure.
In this way it can facilitate communication, similar to any high-level visualisation of a software implementation.
However, the formalised nature of a DSM (as a graph representation) allows for in-depth analysis of the revealed structure.
The case studies have illustrated that this makes it possible to analyse the MDAO architecture by revealing both feedback loops between models and assumptions manifesting in (missing) interfaces.

Firstly, the visual overview of the interdependencies between different models reveals which interfaces are implemented, and which structural assumptions are made in this regard.
For both case studies, missing interfaces have been revealed that contextualise the results produced by the code.
These absent interfaces may often be assumptions that are acknowledged and deemed negligible, but could also be accidental omissions that significantly degrade result accuracy.
Either way, the (x)DSM provides a complete overview for disciplinary experts to validate the set of implemented interfaces and communicate implicit assumptions to non-experts.
In this way, a DSM overview can improve the transparency of complex codes, supporting interpretation of the results.

Secondly, the DSM reveals feedback loops between models, and how they are converged in the MDAO architecture.
For example, the automatically generated PROCESS DSM in \cref{fig:tokamak_collapsed} shows how some feedback is resolved with an MDA solver, and some is handled with a consistency constraint.
Couplings may also be handled using a penalty objective, as \cref{fig:stellarator_xdsm} has revealed.
Furthermore, a DSM can be used to reveal a minimal set of feedback-containing blocks using sequencing algorithms, as applied to \cref{fig:tokamak_expanded}.
One way or another, iteration is required to converge such feedback loops self-consistently.
The MDAO literature provides a reference for different architectures that can be applied~\cite{martinsMultidisciplinaryDesignOptimization2013}.

As iteration drives computational cost, the way feedback loops are handled can significantly affect the computational efficiency of the code.
This is illustrated by the PROCESS case, where a small change in how the models are converged already results in a 20--50\% reduction in model evaluations.
As shown in \cref{subsec:surgery}, in some cases even feedback loops that are slow to converge can be iterated fewer times, as some perturbation vectors for (finite difference) MDA evaluations may affect them less than other loops.
The specific results that are presented may be limited to particular configurations and convergence settings, as PROCESS is quite sensitive to this.
Furthermore, runtime did not decrease significantly compared to default PROCESS, due to the relative importance of the overhead of the MDA convergence checks selected for the purpose of the experiment.\footnote{As Bogaarts et al.~\cite{bogaartsManagingComplexityIntegrated2026b} show, the problem of overhead can be addressed for future frameworks.}
However, the case nonetheless illustrates that architecture can significantly reduce model evaluations.

Towards cases involving many models and higher fidelity models, such architectural considerations can be key to avoiding inefficient execution. 
With higher fidelity models the fraction of execution time spent in model evaluation increases.
When a large number of models are used to represent a complex system, many feedback loops are typically present that may be converged inefficiently if not handled carefully.
If these loops contain expensive models, the impact on runtime can be significant.
How much of this execution time can be saved with architectural considerations depends on the problem, as it may not always be possible to reduce the number of evaluations of the more expensive models.
As discussed for the stellarator case in \cref{subsec:stellopt}, different MDAO architectures may also be suggested from the literature to reduce model evaluations.
Design structure matrices can then be used as a tool to reveal the architecture and identify how the computation can be accelerated.  

It is worth noting that assumptions and feedback loops in MDAO architectures are fundamentally related.
If iteration over a feedback loop is computationally too challenging, the models involved can be decoupled with an assumption.
Conversely, assumptions on (missing) interfaces may indicate hidden design feedback that needs manual verification.
For example, the dependency between coil and plasma optimisation in \cref{fig:stellarator_xdsm} is often decoupled (two-stage approach), assuming the coils can produce the desired surface.
Single-stage optimisation couples these with the $f_{\mathrm{QF}}$ penalty, and alternatively a fully self-consistent treatment may be warranted to increase result accuracy - directly computing $f_{\mathrm{QS}}$ from $\mathbf{B}_{\mathrm{ext}}$, or implementing a strictly enforced constraint.
For highly complex systems, this yields the classic accuracy-cost trade-off: a simple code must make many assumptions, while a thorough code must resolve the complex feedback loops. 
A DSM can reveal both sides of this coin, and help inform what MDAO architecture may be suitable to appropriately solve the design problem.

Though this paper has focused mostly on optimisation, the same principles apply to integrated analysis problems.
Similarly, feedback between models can be important to resolve self-consistently~\cite{meneghiniIntegratedModelingApplications2015}.
Whether an iteration loop is driven by an optimiser or a solver does not matter for the structure that appears in the DSM.
As shown in \cref{fig:tokamak_collapsed}, PROCESS in evaluation mode already contains a complex structure of model couplings and nested solvers.
Architectural considerations impact model evaluations here just as well, as demonstrated with \cref{tab:phaseA_results}.
A difference is that integrated analysis problems typically include time-dependence.
Though MDAO architectures are typically not formulated with explicit time-dependence, similar approaches can be applied to enable optimal control as well (see e.g.~\cite{falckDymosPythonPackage2021}).
Design structure matrices are then similarly applicable to structure couplings between different models.

In future, the DSM approach can be extended to graph-based design of integrated codes and reactors themselves as well.
Using a DSM, one can design the integration structure first to help define the required inputs/outputs of models, essentially constructing an architectural graph before implementation.
A graph-based approach can then enable flexible (re)configuration of MDAO architectures if put at the heart of a software framework~\cite{vangentFormulationIntegrationMDAO2019a}.
Such a framework for stellarator reactor design is currently in development~\cite{bogaartsManagingComplexityIntegrated2026b}.
Furthermore, a graph-based approach can be integrated with model-based systems engineering (MBSE) methods, such as developed in the AGILE project for aerospace~\cite{ciampaAGILEParadigmNext2020,bruggemanMBSEBasedRequirementVerification2022,bussemakerAGILE40Project2023}.
In fact, an MBSE graph can be used as a semantic reference, mapped to the MDAO system with DSMs to reveal further assumptions made in the code~\cite{ruttenIntegratingMBSEMDAO2026}.

\section{Conclusion}
\label{sec:conclusion}
Further digitalisation of fusion reactor design requires more and higher fidelity models to be integrated in design frameworks.
Consequently, the number of ways to distribute and coordinate the involved models, solvers and optimisers increases rapidly.
Having an overview of this MDAO architecture helps to validate the assumptions made in the model couplings and identify inefficient iteration loops.
We illustrate that fusion codes can be expressed as MDAO problems, and (x)DSM methods from the MDAO field can be applied to provide the means to visualise and analyse the structure of internal dependencies.

To that end, DSM methods have been applied to two fusion case studies: stellarator optimisation using SIMSOPT and the PROCESS systems code.
These case studies have demonstrated how analysing the architectural overview provided by a DSM allows:
\begin{enumerate}[leftmargin=*,nosep]
    \item Revealing structural assumptions regarding (missing) interfaces
    \item Revealing feedback loops and how they are resolved
\end{enumerate}
The insight provided by the DSM can then be used to restructure the MDAO architecture to improve result accuracy and/or computational efficiency.
To illustrate potential efficiency gains, we have partitioned the solver (MDA) architecture of PROCESS based on its automatically generated DSM, resulting in a reduction of 20--50\% in model evaluations.
A function-level DSM further indicates that roughly half of the PROCESS functions may not need iteration at all.
This illustrates that MDAO architecture can have a significant impact on computational efficiency.

For future developments of reactor design frameworks including high fidelity codes, architectural considerations are expected to be critical.
With design structure matrices, the structure of these future codes can be designed in a systematic way.

\section*{Declaration of competing interest}
The authors declare that they have no known competing financial interests or personal relationships that could have appeared to influence the work reported in this paper.

\section*{Funding}
This research did not receive any specific grant from funding agencies in the public, commercial, or not-for-profit sectors.

\section*{Declaration of generative AI use}
LLM tools have been used to support the authors in writing this manuscript. Application is limited to software development, literature suggestion, review and editing of the manuscript.
Anthropic Claude Opus and Fable models have been used via Claude Code and claude.ai (Anthropic, https://claude.ai) between March and November 2026.
The authors reviewed and edited the content as needed and take full responsibility for the content of the published article.

\appendix
\setcounter{figure}{0}\setcounter{table}{0} 
\raggedbottom 

\section{PROCESS architecture change full experiment}
\label{app:full_surgery}
This appendix gives the full details of the experiment in \cref{subsec:surgery}, including the evaluation mode experiment summarised in the main text.

\subsection{Experiment cases}
\label{app:surgery_cases}

\Cref{tab:switch_matrix_eval} gives the evaluation mode cases AR to A2, which correspond to the optimisation mode cases BR to B2 in \cref{tab:switch_matrix_opt}, with two exceptions due to the absence of an optimiser.
Firstly, the burn time coupling cannot be delegated to the optimiser, so A1 and A2 instead keep the burn time fixed at the same perturbed initial value.
As this changes the MDA problem rather than its architecture, the effect of the partitioning is measured by the ratio A2/A1 (A2/A0 for \texttt{st}, which has no burn time coupling), whereas in optimisation mode B0 to B2 all solve the same problem.
Secondly, an evaluation mode run consists of a single MDA only, so \texttt{MDA\_Output} is not run at all, and the post-processing models of A2 are executed once after the MDA has converged.
For \texttt{st}, B1 is not run, so that B2 also includes the removal of \texttt{MDA\_Output}.

\begin{table}[H]
    \centering
    \footnotesize
    \setlength{\tabcolsep}{3pt}
    \begin{tabular}{l|cccc}
        \hline
         & AR & A0 & A1 & A2 \\
        \hline
        MDA architecture      & unified & unified & unified & partitioned \\
        MDAs converged on & obj.\,\&\,constr. & all outputs & all outputs & all outputs \\
        \texttt{MDA\_Output}            & - & - & - & - \\
        Burn time coupling & MDA & MDA & fixed & fixed \\
        Models sequenced             & no & no & no & yes \\
        \hline
    \end{tabular}
    \caption{The modifications of PROCESS implemented in each evaluation mode case.}
    \label{tab:switch_matrix_eval}
\end{table}

Evaluation mode initialises the coupling variables from the fixed point, randomly perturbed by up to $\pm 10\%$, and all 25 runs converge in every case.
In optimisation mode, one run starts from the default initial design variables of the optimiser, and the other 24 runs perturb these in the same way (within the specified design variable bounds $\underline{\mathbf{x}}, \overline{\mathbf{x}}$).
Every case, including the reference cases, uses the same set of starting points, so that the cases can be compared run by run in the per-run ratios.

\subsection{Evaluation mode results}
\label{app:surgery_eval}
\Cref{tab:phaseA_results} shows that the control case A0 has similar evaluation counts to the PROCESS reference case (AR), and that fixing the burn time variable (A1) has only limited impact.
As the ratio A2/A1 shows, all modules except M2 converge with significantly fewer model evaluations once the MDA is partitioned (A2), revealing M2 as the bottlenecking feedback loop.
These modules require 20--50\% fewer evaluations, with feedforward and post-processing models running only once.
As a single MDA only requires around five iterations, this amounts to one to three fewer iterations for M1 and M3.

\begin{table}[H]
    \centering
    \footnotesize
    \setlength{\tabcolsep}{3pt}
    \makebox[\columnwidth][c]{%
    \begin{tabular}{l|rrrr|rc}
    \hline
    Module & AR & A0 & A1 & A2 & A2/A1 & A2/A1 med [min, max] \\
    \hline
    \multicolumn{7}{l}{\texttt{tok} ($n = 25$)} \\
    \hline
    M1              & 5.0 & 5.5 & 5.1 & 4.0 & 0.78 & 0.80 [0.67, 1.00] \\
    M2              & 5.0 & 5.5 & 5.1 & 5.2 & 1.01 & 1.00 [1.00, 1.25] \\
    M3              & 5.0 & 5.5 & 5.1 & 3.0 & 0.59 & 0.60 [0.50, 0.75] \\
    Feedforward     & 5.0 & 5.5 & 5.1 & 1 & 0.20 & 0.20 [0.17, 0.25] \\
    Post-processing & 5.0 & 5.5 & 5.1 & 1 & 0.20 & 0.20 [0.17, 0.25] \\
    \hline
    \multicolumn{7}{l}{\texttt{lad} ($n = 25$)} \\
    \hline
    M1              & 5.0 & 5.0 & 4.9 & 4.0 & 0.81 & 0.80 [0.80, 1.00] \\
    M2              & 5.0 & 5.0 & 4.9 & 4.9 & 0.99 & 1.00 [0.80, 1.00] \\
    M3              & 5.0 & 5.0 & 4.9 & 3.0 & 0.61 & 0.60 [0.60, 0.75] \\
    Feedforward     & 5.0 & 5.0 & 4.9 & 1 & 0.20 & 0.20 [0.20, 0.25] \\
    Post-processing & 5.0 & 5.0 & 4.9 & 1 & 0.20 & 0.20 [0.20, 0.25] \\
    \hline
    \multicolumn{7}{l}{\texttt{st} ($n = 25$; ratio A2/A0)} \\
    \hline
    M1              & 4.9 & 5.8 & -- & 4.0 & 0.68 & 0.67 [0.67, 0.80] \\
    M2              & 4.9 & 5.8 & -- & 5.8 & 1.00 & 1.00 [1.00, 1.00] \\
    M3              & 4.9 & 5.8 & -- & 3.0 & 0.51 & 0.50 [0.50, 0.60] \\
    Feedforward     & -- & -- & -- & -- & -- & -- \\
    Post-processing & 4.9 & 5.8 & -- & 1 & 0.17 & 0.17 [0.17, 0.20] \\
    \hline
    \end{tabular}}
    \caption{Average number of model evaluations per module in an evaluation mode run, with the ratio of the averages and the median [min, max] of the per-run ratios.}
    \label{tab:phaseA_results}
\end{table}

\subsection{Optimiser convergence}
\label{app:surgery_iterations}
\Cref{tab:phaseB_iterations} shows the number of optimiser iterations in each case, including the iterations of all optimiser attempts, as PROCESS restarts the optimiser with a different finite difference step size when an attempt fails.
Starts are only accepted when all cases converge. 
Most of the rejected starts also fail for unmodified PROCESS (BR).
For \texttt{tok}, the same three starts fail in every case, and for \texttt{lad}, 13 of the 14 rejected starts already fail in BR, with the remaining start failing in B1 and B2 only.
For \texttt{st}, one start fails in every case, one only in B0 and one only in B2.
The rejection of starts therefore does not seem to favour the modified architecture.
For the accepted starts, optimiser convergence is similar, with the median run in the \texttt{tok} and \texttt{st} configurations converging in the same number of iterations.
The same optimum is reached in all cases except for two starts on \texttt{st} and three on \texttt{lad}.
For \texttt{lad}, these differences all originate from the burn time cut (B0 to B1), while the partitioning (B1 to B2) follows the same optimisation path for all 11 starts.
For \texttt{lad}, the spread in required iterations is quite significant.

\begin{table}[H]
    \centering
    \footnotesize
    \setlength{\tabcolsep}{3pt}
    \makebox[\columnwidth][c]{%
    \begin{tabular}{l|rrrr|rc}
    \hline
    Configuration & BR & B0 & B1 & B2 & B2/B0 & B2/B0 med [min, max] \\
    \hline
    \texttt{tok} ($n = 22$) & 7.8 & 7.8 & 7.8 & 7.8 & 0.99 & 1.00 [0.88, 1.14] \\
    \texttt{lad} ($n = 11$) & 29.8 & 29.8 & 20.9 & 20.9 & 0.70 & 0.81 [0.13, 5.91] \\
    \texttt{st} ($n = 22$) & 31.2 & 25.1 & -- & 24.0 & 0.95 & 1.00 [0.25, 1.36] \\
    \hline
    \end{tabular}}
    \caption{Average number of optimiser iterations per optimisation run, with ratios as in \cref{tab:phaseB_results}.}
    \label{tab:phaseB_iterations}
\end{table}

\subsection{Wall-clock time}
\label{app:wallclock time}

The wall-clock time per optimisation run is shown in \cref{tab:wallclock_tok,tab:wallclock_lad,tab:wallclock_st}, for the same starts as \cref{tab:phaseB_results}.
The times were measured with timers in the experiment harness (v0.5.0), with three runs in parallel on one workstation, and should be considered indicative only.
The time spent in the models of each module follows the number of model evaluations in \cref{tab:phaseB_results} closely.
Most of the overhead is introduced by the convergence criterion of the control case: checking the convergence of all model outputs takes 0.1--0.3\,s per run in BR, but 10--40\,s in B0, which is about two thirds of the time spent in the models themselves.
As this check scales with the number of MDA iterations, it is reduced by the modified architecture as well, so that B2 runs 22--45\% faster than B0.
Compared to unmodified PROCESS (BR), however, the additional convergence check mostly cancels this speedup: B2 is 11--16\% faster than BR for \texttt{lad} and \texttt{st}, and 25\% slower for \texttt{tok}.

\begin{table}[H]
\centering
\footnotesize
\setlength{\tabcolsep}{2pt}
\begin{tabular}{l|cccc|cc}
\hline
Row & BR & B0 & B1 & B2 & B2/B0 & \begin{tabular}[c]{@{}c@{}}B2/B0\\ {}med [min, max]\end{tabular} \\
\hline
\multicolumn{7}{l}{\texttt{tok} ($n = 22$)} \\
\hline
M1                        & 6.69 & 7.54 & 7.41 & 5.06 & 0.67 & 0.69 [0.53, 0.84] \\
M2                        & 6.52 & 7.25 & 7.14 & 6.09 & 0.84 & 0.87 [0.66, 1.08] \\
M3                        & 1.37 & 1.52 & 1.49 & 1.11 & 0.73 & 0.73 [0.57, 0.95] \\
Feedforward               & 0.02 & 0.02 & 0.02 & 0.01 & 0.34 & 0.35 [0.24, 0.46] \\
Post-processing           & 0.39 & 0.42 & 0.42 & 0.00 & 0.00 & 0.00 [0.00, 0.00] \\
MDA convergence           & 0.08 & 10.63 & 10.48 & 8.22 & 0.77 & 0.79 [0.63, 1.01] \\
dispatch                  & 0.23 & 0.30 & 0.31 & 0.55 & 1.80 & 1.89 [1.25, 2.38] \\
obj.\,\&\,constr.         & 0.33 & 0.14 & 0.14 & 0.14 & 1.02 & 1.01 [0.52, 2.01] \\
optimiser own time        & 0.15 & 0.11 & 0.10 & 0.10 & 0.92 & 0.95 [0.45, 2.56] \\
fixed per run             & 5.11 & 5.28 & 4.59 & 4.62 & 0.88 & 0.88 [0.73, 1.02] \\
unattributed residual     & 0.03 & 0.11 & 0.11 & 0.15 & 1.35 & 1.42 [0.92, 1.96] \\
\hline
Total                     & 20.92 & 33.33 & 32.21 & 26.05 & 0.78 & 0.80 [0.62, 0.94] \\
\hline
\end{tabular}
    \caption{
        Wall-clock time per optimisation run (s) for the large tokamak (\texttt{tok}), with ratios as in \cref{tab:phaseB_results}.
        Module rows give the time spent in the models of each module, \textit{MDA convergence} the convergence checks of the MDAs, \textit{dispatch} the remaining MDA overhead, \textit{obj.\,\&\,constr.} the evaluation of the objective and constraints, \textit{optimiser own time} the solve time outside model evaluations, and \textit{fixed per run} start-up, input and output; the \textit{unattributed residual} closes the measured total.
        }
    \label{tab:wallclock_tok}
\end{table}

\begin{table}[H]
\centering
\footnotesize
\setlength{\tabcolsep}{2pt}
\begin{tabular}{l|cccc|cc}
\hline
Row & BR & B0 & B1 & B2 & B2/B0 & \begin{tabular}[c]{@{}c@{}}B2/B0\\ {}med [min, max]\end{tabular} \\
\hline
\multicolumn{7}{l}{\texttt{lad} ($n = 11$)} \\
\hline
M1                        & 27.62 & 27.17 & 18.54 & 12.68 & 0.47 & 0.57 [0.08, 4.11] \\
M2                        & 28.65 & 27.88 & 19.03 & 16.43 & 0.59 & 0.71 [0.10, 5.22] \\
M3                        & 5.89 & 5.60 & 3.81 & 3.02 & 0.54 & 0.65 [0.09, 4.89] \\
Feedforward               & 0.07 & 0.07 & 0.04 & 0.02 & 0.25 & 0.30 [0.04, 2.30] \\
Post-processing           & 1.64 & 1.53 & 1.03 & 0.00 & 0.00 & 0.00 [0.00, 0.00] \\
MDA convergence           & 0.32 & 40.16 & 27.25 & 21.83 & 0.54 & 0.65 [0.09, 4.98] \\
dispatch                  & 0.97 & 1.08 & 0.74 & 1.37 & 1.27 & 1.60 [0.22, 11.59] \\
obj.\,\&\,constr.         & 1.42 & 0.50 & 0.36 & 0.37 & 0.75 & 0.75 [0.17, 7.01] \\
optimiser own time        & 0.38 & 0.31 & 0.23 & 0.24 & 0.77 & 0.78 [0.13, 6.36] \\
fixed per run             & 5.46 & 5.19 & 4.44 & 4.48 & 0.86 & 0.86 [0.80, 0.99] \\
unattributed residual     & 0.23 & 0.47 & 0.31 & 0.42 & 0.88 & 1.13 [0.14, 9.36] \\
\hline
Total                     & 72.64 & 109.97 & 75.78 & 60.86 & 0.55 & 0.67 [0.10, 4.34] \\
\hline
\end{tabular}
    \caption{Wall-clock time per optimisation run (s) for the low aspect ratio DEMO (\texttt{lad}), as in \cref{tab:wallclock_tok}.}
    \label{tab:wallclock_lad}
\end{table}

\begin{table}[H]
\centering
\footnotesize
\setlength{\tabcolsep}{2pt}
\begin{tabular}{l|cccc|cc}
\hline
Row & BR & B0 & B1 & B2 & B2/B0 & \begin{tabular}[c]{@{}c@{}}B2/B0\\ {}med [min, max]\end{tabular} \\
\hline
\multicolumn{7}{l}{\texttt{st} ($n = 22$)} \\
\hline
M1                        & 20.37 & 17.64 & -- & 11.09 & 0.63 & 0.69 [0.16, 0.91] \\
M2                        & 19.93 & 16.89 & -- & 11.10 & 0.66 & 0.69 [0.17, 1.00] \\
M3                        & 5.01 & 4.17 & -- & 2.68 & 0.64 & 0.70 [0.16, 0.96] \\
Feedforward               & -- & -- & -- & -- & -- & -- \\
Post-processing           & 1.17 & 0.98 & -- & 0.00 & 0.00 & 0.00 [0.00, 0.00] \\
MDA convergence           & 0.22 & 24.92 & -- & 16.58 & 0.67 & 0.71 [0.17, 1.01] \\
dispatch                  & 0.62 & 0.64 & -- & 1.00 & 1.57 & 1.61 [0.39, 2.49] \\
obj.\,\&\,constr.         & 0.81 & 0.25 & -- & 0.22 & 0.86 & 0.89 [0.22, 1.31] \\
optimiser own time        & 0.29 & 0.24 & -- & 0.24 & 0.99 & 0.99 [0.21, 1.84] \\
fixed per run             & 5.19 & 5.17 & -- & 4.60 & 0.89 & 0.89 [0.76, 1.03] \\
unattributed residual     & 0.13 & 0.30 & -- & 0.34 & 1.12 & 1.11 [0.24, 1.84] \\
\hline
Total                     & 53.73 & 71.20 & -- & 47.83 & 0.67 & 0.72 [0.19, 0.98] \\
\hline
\end{tabular}
    \caption{Wall-clock time per optimisation run (s) for the spherical tokamak (\texttt{st}), as in \cref{tab:wallclock_tok}.}
    \label{tab:wallclock_st}
\end{table}

\bibliographystyle{elsarticle-num}
\bibliography{xDSM_for_fusion,own_papers}

\end{document}